\documentclass[aps,preprint,eqsecnum,showpacs,preprintnumbers,byrevtex,nofootinbib]{revtex4}
\usepackage{amssymb,amsmath,latexsym}
\usepackage{hyperref}
\newcommand{\fracmy}[2]{\frac{#1}{#2}\:}
\newcommand{\sps}{\hspace{0.13cm}}
\newcommand{\dsps}{\hspace{0.26cm}}
\newcommand{\tsps}{\hspace{0.39cm}}

\newcommand{\dimi}[1]{~\hspace{-0.1cm} ^{(#1)} \hspace{-0.1cm}}
\newcommand{\esi}[1]{\stackrel{\hspace{-0.4cm}*}{#1}}
\newcommand{\es}[1]{\stackrel{\hspace{0.1cm}*}{#1}}
\newcommand{\ecnabla}[1]{\nabla_{\hspace{-0.1cm} #1 \hspace{0.05cm}}}
\newcommand{\enabla}[1]{\stackrel{\hspace{-0.1cm}*}{\nabla_{\hspace{-0.1cm} #1 \hspace{0.05cm}}}}
\newcommand{\uenabla}[1]{\stackrel{\hspace{-0.1cm}*}{\nabla^{\hspace{0.0cm} #1 \hspace{0.05cm}}}}
\begin{document}

\title{Effective Equations on the 3-Brane World from Type IIB String}

\author{Denis Gon\c{t}a}

\email[e-mail: ]{d.gonta@pobox.spbu.ru}

\affiliation{\mbox{Department of Physics, St. Petersburg State
University,} \\ St. Petersburg, 198504, Russia}

\begin{abstract}
The effective field equations on a 3-brane are established
considering the massless bosonic sector of the type $IIB$ string
compactified on $S^5$. The covariant embedding formalism in a
space endowed with $Z_2$-symmetry is applied. Recently the
derivation of effective equations on the 3-brane, where only
gravity penetrates in the bulk has been performed by Shiromizu,
Maeda, and Sasaki \cite{br1a}. We extend this analysis to the
situation when the bulk contains a set of fields given by the type
$IIB$ string. The notion of the Einstein-Cartan space is
considered in order to avoid extra suppositions about the
embedding of these fields. The interactions between the brane and
the bulk fields are understood in a purely geometric way, which
fixes the form of these interactions. Finally, we present the
dynamically equivalent effective equations have expressed
completely in Riemannian terms and make conclusions.
\end{abstract}

 \preprint{gr-qc/0503009}
\maketitle

\setcounter{section}{0}

\section*{\normalsize INTRODUCTION AND MOTIVATIONS}

Since the papers of Kaluza and Klein \cite{br0} it has been
suggested a possibility that there exist extra dimensions beyond
those of Minkowski space-time. In recent years, ideas of extra
dimensions become much more compelling. According to \cite{br6,
br1a, br1b} it has been understood that additional dimensions
could have a quite distinct nature from those of Kaluza and Klein.
In other words, ordinary matter would confine to our 4-dimensional
world while gravity would penetrate in the extended space (bulk).
An intriguing motivation for considering these models comes from
string/M-theory. For instance, the 10-dimensional type $IIB$
theory can admit brane worlds solutions \cite{str19, str20}. The
type $IIB$ theory have a rich massless spectrum. Hence this theory
illustrates an enlarged complexity due to the supplementary fields
in addition to gravity and even scalar fields (see
\cite{str20,br4,str22} for previous work). This situation is more
curious since it seems unlikely that only gravity can penetrate in
extra dimensions. In this paper, considering the massless bosonic
sector of the type $IIB$ theory have compactified on a sphere
$S^5$ we derive effective field equations on the 3-brane in a
5-dimensional bulk endowed with $Z_2$-symmetry.

Our approach to this problem is based on the covariant embedding
formalism, which gives a coordinate independent derivation of the
dynamics on the brane. In a typical brane worlds setup, where only
gravity penetrates in the bulk, the derivation of effective field
equations was performed in \cite{br1a,br1b}. We extend this
analysis to the situation, which assumes the bulk to contain a set
of fields given by the type $IIB$ theory. The problem that arises
is to find an appropriate description of the embedding procedure
in presence of additional bulk fields in order to avoid extra
suppositions about the embedding of these fields. In this paper,
we consider one possible candidate for such description, which is
based on the notion of the Einstein-Cartan space. Roughly speaking
this space is a generalization of the Riemannian space by
considering the torsion tensor \cite{tors2,misc3}. The status of
torsion in the framework of 4-dimensional gravity and cosmology
remains open today \cite{tors2, tors4, tors6, tors7, tors8}.
However such spaces have recently become relevant due to the
relationship to many extended theories like superstrings,
supergravity, and e.t.c. \cite{str2, str3, str5, str6, tors4,
tors5}. Hence it has been expected that the embedding formalism
developed for the Einstein-Cartan space can give an unified and
transparent understanding of the dynamics on the brane. At the
same time we stress that one can express the resulting effective
equations in several dynamically equivalent forms. Thus to
consider the Riemannian form supplied with extra terms of
non-Riemannian origin.

The organization of the article is as follows. In the first
section we start with basic facts about the type $IIB$ theory and
the Einstein-Cartan space. We establish the connection between the
torsion and a combination of bulk fields. In the second section
taking account of the covariant embedding formalism constructed in
the Einstein-Cartan space (see Appendix \ref{app1}) we derive the
effective field equations on the brane. Finally, in the last
section we make conclusions and discuss the results.

\section{Model Building}

In fact, the strings can be formulated in curved spaces in
presence of massless background fields. In this case, conformal
invariance conditions for a closed bosonic string become
equivalent to equations of motion for the following background
fields: the antisymmetric tensor field $B_{AB}$, the dilaton field
$\Phi$, and the metric $G_{AB}$\footnote{Where the uppercase Latin
indices $A,B = 0,\ldots,D-1$.}. These equations can be derived
from the following action \cite{str1,str11}\footnote{$A_{[a_1,
\ldots, a_p]} \equiv \fracmy{1}{p!}\sum (-1)^\pi A_{a_{\pi 1}
\ldots \: a_{\pi p}}.$}
\begin{equation}\label{string-eff0}
S_{b} = \frac{1}{2k^2_{26}}\int{d^{26}X \sqrt{|G|} \: e^{-2\Phi}
\left( \es{R} - \fracmy{1}{12} H_{ABC}H^{ABC} + 4
\partial_A\Phi \partial^A\Phi \right)}; \: H_{ABC} \equiv 3\partial_{[A}
B_{BC]},
\end{equation}
where $\es{R}$ is the curvature scalar computed from the metric
$G_{AB}$ and $k^2$ is the gravitational constant. On the other
hand, the massless bosonic part of the action of superstring
theories can be expressed in the form
\begin{equation}\label{link0}
S_{superstring}  = S_{universal} + S_{model} + S_{interactions},
\end{equation}
where $S_{universal}$ does not depend on which of superstring
theories we consider and it has the form (\ref{string-eff0}) taken
at $D = 10$. The $S_{model}$ part depends on the superstring
theory we consider and it contains the Ramond fields: $\{A_C,
W_{ABC}\}$ (type $IIA$ string) or $\{\Psi, A_{BC}, W_{ABCD}\}$
(type $IIB$ string). The last term in (\ref{link0}) is the
Chern-Simons like term.

Motivated by brane worlds ideology and 4-dimensional cosmology the
bulk should be the \textit{almost-AdS$_5$} space, which represents
a space that is \textit{AdS$_5$} like on a large scale but allows
for generic inhomogeneities on a small scale. Hence a suitable
compactification of a specified superstring theory should be
considered. Such situation is possible in the framework of the
type $IIB$ theory compactified on $S^5$. Incidentally, the
massless bosonic sector of the type $IIB$ theory can not exactly
be described by a covariant action \cite{str17}, but the covariant
equations of motion exist \cite{str18}. This situation arises
whenever the dimensions of space-time is $D=4n+2$. In this case,
the $2n+1$ index field strength may be self-dual in locally
Minkowski spaces. Hence the set of equations should contain the
self-duality condition. Nevertheless these equations can be
derived from the following action \cite{str11}
\begin{eqnarray}\label{string-eff1}
S_\sigma = \frac{1}{2k^2_{10}}\int d^{10}X \sqrt{|G|}
\hspace{-0.4cm} && \left[ e^{-2\widehat{\Phi}} \left(
\dimi{10}\es{R} - \fracmy{1}{12} H_{ABC}H^{ABC} + 4
\partial_A\widehat{\Phi} \partial^A\widehat{\Phi} \right) \right.
\notag \\ && - \left. \fracmy{1}{2} \partial_A \Psi \partial^A
\Psi - \fracmy{1}{12} \widetilde{F}_{ABC}\widetilde{F}^{ABC} -
\fracmy{1}{480} \widetilde{Y}_{ABCDE}\widetilde{Y}^{ABCDE}
\right];
\end{eqnarray}
\begin{eqnarray}
H_{ABC} &=& 3\partial_{[A}B_{BC]}, \quad F_{ABC} =
3\partial_{[A}A_{BC]}, \quad \widetilde{F}_{ABC} = F_{ABC} - \Psi
H_{ABC}; \notag \\ Y_{ABCDE} &=& 5\partial_{[A}W_{BCDE]}, \quad
\widetilde{Y}_{ABCDE} = Y_{ABCDE} - 5 A_{[AB}H_{CDE]} + 5
B_{[AB}F_{CDE]}; \notag
\end{eqnarray}
except for a term $\widetilde{Y}^2$ in one of equations and for
the self-duality condition $\widetilde{Y}_{ABCDE} =
*\widetilde{Y}_{ABCDE}$. Taking account of these equations the
compactification on $S^5$ proposed in \cite{str12} has widely used
\cite{br7}. On the other hand, the action (\ref{string-eff1}) has
derived in the $\sigma$-model frame. Such frame should be not
confused with the coordinate frame in general relativity. The
string theory is insensitive to local redefinitions of background
fields \cite{str14}. Therefore in order to obtain a standard
normalized Einstein action (in the 5-dimensional case) one should
redefine the metric as follows
\begin{equation}\label{frame}
G_{AB} = e^{\frac{4\Phi}{D-2}} \: \widetilde{G}_{AB}, \quad
\text{where} \quad \widehat{\Phi} \equiv \Phi +
\fracmy{5}{8}\widetilde{\Phi}.
\end{equation}
It can straightforwardly be checked that the action
(\ref{string-eff1}) becomes
\begin{eqnarray}\label{string-eff2}
S_E = \frac{1}{2k^2_{10}}\int d^{10}X \sqrt{|\widetilde{G}|} \:
e^{-\fracmy{5}{4}\widetilde{\Phi}} \hspace{-0.4cm} && \left[
\es{\widetilde{R}} - \fracmy{e^{-\Phi}}{12}H^2 - \fracmy{1}{2}
\partial_A \Phi \partial^A \Phi - \fracmy{5}{8}\partial_A \Phi
\partial^A \widetilde{\Phi} + \fracmy{25}{16} \partial_A \widetilde{\Phi} \partial^A
\widetilde{\Phi}  \right. \notag \\ &&  \left. - \fracmy{e^{2\Phi
+ \fracmy{5}{4}\widetilde{\Phi}}}{2} \partial_A \Psi \partial^A
\Psi - \fracmy{e^{\Phi + \fracmy{5}{4}\widetilde{\Phi}}}{12}
\widetilde{F}^2 - \fracmy{e^{\fracmy{5}{4}\widetilde{\Phi}}}{480}
\widetilde{Y}^2 \right].
\end{eqnarray}
By the frame (\ref{frame}) we have increased the number of
independent fields by one. This introduces an additional
invariance and implies that one of equations derived from this
action has kinematically related to the remaining set of
equations. Hence this equation can be eliminated. Notice that by
the compactification ansatz given below this new field plays the
role of moduli. The equations of motion obtained from those
considered in \cite{str12} by the redefinition (\ref{frame}) can
realdily be derived from the action (\ref{string-eff2}) taking
account of the term $\widetilde{Y}^2$ and the self-duality
condition. These equations are listed in the Appendix \ref{app3}
[see (\ref{eqs-motion})].

Further, taking into account the action (\ref{string-eff2})
instead of (\ref{string-eff1}) we apply the ansatz \cite{str12}
\begin{subequations}\label{ansatz}
\begin{eqnarray}
ds^2_{10} = \widetilde{G}_{AB}dX^A dX^B = g_{ab}(x)dx^a dx^b +
e^{\rho(x)/2} g_{\underline{ab}}(\theta) d\theta^{\underline{a}}
d\theta^{\underline{b}}; \label{ansatz1} \hspace{3cm} \\ \Phi(X) =
\phi(x), \: \widetilde{\Phi}(X) = \rho(x), \: B_{AB}(X) =
B_{ab}(x) \delta_{AB}^{ab}, \: \Psi(X) = \chi(x), \: A_{AB}(X) =
A_{ab}(x) \delta_{AB}^{ab}, \notag
\\ W_{ABCD}(X) = W_{abcd}(x) \delta_{ABCD}^{abcd}
+ \varrho \: \widetilde{W}_{\underline{abcd}}(\theta)
\delta_{ABCD}^{\underline{abcd}} \quad \text{such that} \quad 5
\partial_{[\underline{a}} \widetilde{W}_{\underline{bcde}]} =
\eta_{\underline{abcde}}, \label{ansatz2} \hspace{1cm}
\end{eqnarray}
\end{subequations}
where $x^a$ are 5-dimensional coordinates,
$\theta^{\underline{a}}$ parametrizes the sphere $S^5$, $\delta_{A
\ldots B}^{C \ldots D} \equiv \delta_A^C \ldots \delta_B^D$,
$\eta_{\underline{abcde}}$ is the volume form defined on $S^5$,
and $\varrho$ is the Freund-Rubin parameter. We stress that in
order to retain the direct relationship between the different
dimensional theories one should have the truncation to the
low-dimensional subset of fields be a consistent one. In other
words, the solutions of 5-dimensional equations of motion should
give the solutions of original 10-dimensional theory. It has been
widely believed \cite{str19,str20} that every sphere
compactification is a consistent one. However according to
\cite{str13,str22} the situation seems to be more complicate.
Fortunately, the authors of \cite{str13} have shown the strong
consistency of \textit{$AdS_5 \times S^5$} compactfication. Hence
we suppose that the compactification based on
\textit{almost}-$AdS_5 \times S^5$ can admit a consistent
truncation too. The structure of the \textit{almost-AdS$_5$} space
is understood as follows. Suppose that all the background fields
except $\widetilde{G}_{AB}$ and $W_{ABCD}$ can be neglected; then
equations of motion can be solved. The solution describes an
infinite $D3$-brane \cite{str15}, which metric close to the
horizon is given by
\begin{equation}\label{d3-brane}
ds^2_{10} = \fracmy{r^2}{L^2} \left( -dt^2 + \sum_{i=1}^{3} dx^i
dx^i \right) + \fracmy{L^2}{r^2} dr^2 + L^2 d\Omega^2_5; \quad L =
const. \notag
\end{equation}
This metric describes exactly the \textit{$AdS_5 \times S^5$}
space. In general, the existence of the \textit{almost-AdS$_5$}
background in the type $IIB$ theory should be proven. However
taking account of above arguments we conjecture it.

After cumbersome calculations (similar to \cite{str12}) one yields
the 5-dimensional equations of motion (\ref{eqs-motion-low})
listed in the Appendix {\ref{app3}}, which can be derived from the
following action\footnote{Where the lowercase Latin indices $a,b =
0,1,2,3,4$; the Greek indices $\mu,\nu = 0,1,2,3$; and $x_4 \equiv
y$.}
\begin{eqnarray}\label{act0}
S_0 = \frac{1}{2k^2}\int d^{5}x \sqrt{|g|} \hspace{-0.4cm} &&
\left[\es{R} - \fracmy{1}{2} \left( \partial_a\phi
\partial^a\phi + \fracmy{5}{8} \partial_a\rho \partial^a\rho +
\fracmy{5}{4} \partial_a\phi \partial^a\rho + e^{2\phi +
\fracmy{5}{4}\rho}\partial_a\chi \partial^a\chi \right) \right.
\notag \\ && - \left. \fracmy{1}{12} \left( e^{\phi +
\fracmy{5}{4}\rho} \widetilde{F}^2 + e^{-\phi}H^2 \right) -
\fracmy{e^{\fracmy{5}{4}\rho}}{480} \widetilde{Y}^2 +
e^{-\fracmy{1}{2}\rho} \alpha \right];
\end{eqnarray}
\begin{eqnarray}
F_{abc} &=& 3\partial_{[a}A_{bc]}, \quad H_{abc} =
3\partial_{[a}B_{bc]}, \quad \widetilde{F}_{abc} = F_{abc} - \chi
H_{abc}, \notag \\ Y_{abcde} &=& 5\partial_{[a}W_{bcde]}, \quad
\widetilde{Y}_{abcde} = Y_{abcde} - 5 A_{[ab}H_{cde]} + 5
B_{[ab}F_{cde]}, \notag
\end{eqnarray}
except for the terms $\widetilde{Y}^2$ and for consequences of the
self-duality condition in the form
\begin{eqnarray}\label{some8}
\varrho &=& \sqrt{|g|} \: e^{\fracmy{5}{4}\rho} \:
\widetilde{Y}^{01234} = - \fracmy{1}{\sqrt{|g|}} \:
e^{-\fracmy{5}{4}\rho} \: \widetilde{Y}_{01234} = const. \quad
\text{hence} \notag \\ \widetilde{Y}^{abcde} &=& -\sqrt{|g|} \:
\eta^{abcde} \: \widetilde{Y}^{01234} = -\varrho \: \eta^{abcde}
\: e^{-\fracmy{5}{4}\rho}, \notag \\ \widetilde{Y}_{abcde} &=&
\fracmy{1}{\sqrt{|g|}} \: \eta_{abcde} \: \widetilde{Y}_{01234} =
- \varrho \: \eta_{abcde} \: e^{\fracmy{5}{4}\rho}, \notag \\
\text{and} && \widetilde{Y}_{abcde} \widetilde{Y}^{abcde} = -
\widetilde{Y}_{\underline{abcde}}
\widetilde{Y}^{\underline{abcde}} = - 5! \: \varrho^2.
\end{eqnarray}
Notice that $\alpha = const.$ is the scalar curvature of $S^5$,
the 5-dimensional scalar curvature $\es{R}$ has computed from
$g_{ab}$, and $\eta_{abcde}$ is the volume form defined on the
\textit{almost-AdS$_5$} space. Finally, $k$ is the 5-dimensional
coupling constant related to the string scale as follows
\begin{equation}
\fracmy{1}{k^2} = \frac{\text{volume of }S^5}{k^2_{10}}, \notag
\end{equation}
such that $k^2 = 8\pi M_p^{-3}$, where $M_p$ is the fundamental
5-dimensional Planck mass.

Actually one particular combination of Neveu-Schwarz and Ramond
fields can safely be interpreted as the bulk torsion in the
framework of Einstein-Cartan (EC) space (see \cite{str2, str3,
str5, str6, tors7, br2} for related work). Indeed, recall the
following definitions \cite{tors2,misc3}
\begin{eqnarray}
\Gamma^a_{\sps bc} &=& \Gamma^a_{\sps (bc)} + \Gamma^a_{\sps [bc]}
= \esi{\Gamma^a_{\sps bc}} + \left(\Upsilon^a_{\sps bc} +
\Upsilon_{bc}^{\sps \sps a} + \Upsilon_{cb}^{\sps \sps a}\right)
\equiv \: \esi{\Gamma^a_{\sps bc}} - \: Q^{\dsps a}_{cb}
\label{contdef} \quad \text{and} \\ \ecnabla{c} g_{ab} &\equiv&
\partial_c g_{ab} - \Gamma^d_{\sps ac}g_{db} - \Gamma^d_{\sps
bc}g_{ad} = 0; \; \enabla{c} g_{ab} \equiv
\partial_c g_{ab} - \esi{\Gamma^d_{\sps ac}}g_{db} -
\esi{\Gamma^d_{\sps bc}}g_{ad} = 0, \quad \notag
\end{eqnarray}
where $\esi{\Gamma_{abc}}$ is the Christoffel connection, $Q_{abc}
= - Q_{acb} $ is the contortion, and $\Upsilon_{abc} \equiv
\Gamma_{a[bc]}$ is the Cartan's torsion tensor, a purely affine
quantity. If torsion vanishes, we naturally recover the Riemannian
space. The contortion can covariantly be split into irreducible
parts
\begin{equation}\label{irrparts}
Q_{abc} =\fracmy{1}{(D-1)}(Q_{b}g_{ac} - Q_{c}g_{ab}) + L_{abc} +
Q_{[abc]},
\end{equation}
where $Q_a \equiv Q^b_{\sps ab}$ is the trace, $L_{abc}$ is a
tensor constrained by $L^a_{\sps ba} = 0, \; L_{abc}\eta^{abc \:
i_1 \ldots i_{D-3}} = 0$, and $\eta_{i_1 \ldots i_D}$ is the
volume form.

Now instead of the action (\ref{act0}) defined in the Riemannian
bulk consider the following action defined in the EC bulk
\begin{eqnarray}\label{act1}
S = \frac{1}{2k^2}\int d^{5}x \sqrt{|g|} \hspace{-0.4cm} &&
\left[R - \fracmy{1}{2} \left( \partial_a\phi
\partial^a\phi + \fracmy{5}{8} \partial_a\rho \partial^a\rho +
\fracmy{5}{4} \partial_a\phi \partial^a\rho + e^{2\phi +
\fracmy{5}{4}\rho}\partial_a\chi \partial^a\chi \right) \right.
\notag \\ && - \left. \fracmy{1}{12} e^{-\phi} H^2 -
\fracmy{e^{\fracmy{5}{4}\rho}}{480} \widetilde{Y}^2 +
e^{-\fracmy{1}{2}\rho} \alpha \right],
\end{eqnarray}
where the scalar curvature\footnote{$R^a_{\sps bcd} =
\partial_c \Gamma^a_{\sps bd} -
\partial_d \Gamma^a_{\sps bc} + \Gamma^a_{\sps fc} \Gamma^f_{\sps bd}
- \Gamma^a_{\sps fd} \Gamma^f_{\sps bc}$, $R_{ab} = R^c_{\sps
acb}$, $R = g^{ab}R_{ab}$.} is constructed from $\{g_{ab},
Q_{abc}\}$. Using (\ref{contdef}) and (\ref{irrparts}) one can
readily compare equations of motion derived from (\ref{act0}) and
(\ref{act1}). The conclusion is that these sets of equations
coincide whenever
\begin{equation}\label{rule1}
Q_{[abc]} =  \fracmy{1}{2\sqrt{3}} e^{\fracmy{1}{2}\left(\phi +
\fracmy{5}{4}\rho\right)} \widetilde{F}_{abc} \equiv \Theta_{abc},
\quad Q_a = 0, \quad L_{abc} = 0.
\end{equation}
In other words, the theory given by (\ref{act0}) in the Riemannian
space is dynamically equivalent to the theory given by
(\ref{act1}) in the EC space. A combination of Neveu-Schwarz and
Ramond fields is identified with the $Q_{[abc]}$ part of the bulk
contortion. Throughout the paper we shall deal with the action
(\ref{act1}), which underlie the idea of EC space. Notice that in
the EC space one should distinguish two classes of extremal
curves. The autoparallel curves and the geodesic curves. However
these two classes coincide whenever the contortion is totally
antisymmetric \cite{tors2}. Incidentally, this is exactly the case
considered in this paper.

In our brane world scenario the 4-dimensional world $(q_{\mu \nu},
\dimi{4} \Theta_{\alpha \beta \gamma}, \psi_A)$\footnote{Further,
the uppercase Latin indices denote the matter sector of the
brane.} is understood as a brane in the 5-dimensional EC bulk
$(g_{ab}, \Theta_{abc}, W_{abcd})$ characterized by the contortion
(\ref{rule1}). The brane matter sector $L(\psi_A,
\dimi{4}\ecnabla{\mu}\psi_A, q_{\mu \nu})$ consists of a
continuous spinning medium $\psi_A$ minimally coupled to the
induced metric $q_{\mu \nu}$ and the induced contortion $\dimi{4}
\Theta_{\lambda \mu \nu}$. The spinning medium (i.e., Maxwell and
Yang-Mills fields, the Proca field, the Dirac field, and the
Weyssenhoff-Raabe spin fluid) is a matter, which possess the
intrinsic spin, that is the irreducible spin of elementary
particles (see \cite{tors4, tors6, tors8}, where the Lagrangian
description in the EC spaces is given). Taking account of minimal
coupling the brane Lagrangian can be decomposed as follows
\begin{equation}\label{emt-decomp}
L(\psi_A, \dimi{4}\ecnabla{\mu}\psi_A, q_{\mu \nu}) =
\es{L}(\psi_A, \dimi{4}\enabla{\mu}\psi_A, q_{\mu \nu}) +
\widetilde{L}(q_{\mu \nu}, \dimi{4} \Theta_{\alpha \beta \gamma},
\psi_A),
\end{equation}
where $\es{L} \: \equiv L|_{\dimi{4} \: \Theta_{\alpha \beta
\gamma} \rightarrow 0}$ and $\widetilde{L}$ describes the
interaction with the bulk fields $\dimi{4} \: \Theta_{\alpha \beta
\gamma}$.

The equations of motion that describe the compactified type $IIB$
theory in presence of brane are listed in Appendix \ref{app2} [see
(\ref{f-eqs})]. Nevertheless we observe that these equations can
be derived from the following action
\begin{eqnarray}\label{action}
&& \hspace{5cm} S_{total} = S + S_{c} + S_{br} + S_{cbr};
\\ S &=& \frac{1}{2k^2}\int d^{5}x \sqrt{|g|} \left[R - \fracmy{1}{2}
\left( \partial_a\phi \partial^a\phi + \fracmy{5}{8}
\partial_a\rho \partial^a\rho + \fracmy{5}{4} \partial_a\phi
\partial^a\rho + e^{2\phi + \fracmy{5}{4}\rho}\partial_a\chi
\partial^a\chi \right) \right. \notag \\ && \hspace{3cm} - \left.
\fracmy{1}{12} e^{-\phi} H_{abc}H^{abc} -
\fracmy{e^{\fracmy{5}{4}\rho}}{480} \widetilde{Y}_{abcde}
\widetilde{Y}^{abcde} + e^{-\fracmy{1}{2}\rho} \alpha \right],
\notag \\ S_{c} &=& \int{d^{5}x \sqrt{|g|} \left(- \Lambda
\right)}, \quad S_{cbr} = \int{d^{4}x \sqrt{|q|} \left( - \lambda
\right)}, \quad S_{br} = \int{d^{4}x \sqrt{|q|} \: L(q_{\mu \nu},
\psi_A, \dimi{4}\ecnabla{\mu}\psi_A)}, \notag
\end{eqnarray}
except for the terms $\widetilde{Y}^2$ and for consequences of the
self-duality condition [see (\ref{some9})]. Also, $\Lambda \:
(<0)$ is the bulk cosmological constant and $\lambda \: (>0)$ is
the intrinsic brane tension.

The appearance of the negative bulk cosmological constant in the
action (\ref{action}) does not follow directly from the type $IIB$
theory. Recall that the structure of \textit{almost-AdS$_5$} space
supposes the existence of the negative cosmological constant,
which shows up on a large scale. Hence the introduction of one
additional negative constant does not change the picture. The
$Z_2$-symmetry of the fields $(g_{ab}, \rho, \phi, \chi, A_{ab},
B_{ab})$ is also an interesting topic. In our case, the
compactification procedure given in the  Appendix \ref{app3} does
not imposes some special symmetries on the fields $(g_{ab}, \rho,
\phi, \chi, A_{ab}, B_{ab})$ but supplies every point of the bulk
with a 5-sphere scaled by $\rho(x)$ [see (\ref{ansatz1})]. To make
the $Z_2$-symmetry be compatible with the sphere compactification
one must combine the flip in $y$ with an orientation-reversing of
spheres \cite{str19}. In other words, we must operate with an
extended notion of the $Z_2$-transformation that means: $y
\rightarrow -y$ as well $\varrho \rightarrow -\varrho$. At the
same time the bulk metric $g_{ab}$ can always be taken
$Z_2$-symmetric (in the extended sense) by considering the patches
of \textit{almost-AdS$_5$} space glued together along the brane
world volume in a $Z_2$-symmetric manner. The situation with
$(\rho, \phi, \chi, A_{ab}, B_{ab})$ is less optimistic.
Nevertheless it is believed \cite{str19} that in the framework of
the type $IIB$ theory exists a mode-locking mechanism that
projects these fields into a $Z_2$-invariant subspace. Therefore
throughout the paper we shall consider the fields $(g_{ab}, \rho,
\phi, \chi, A_{ab}, B_{ab})$ being symmetric in the extended sense
of the $Z_2$-transformation.

\section{Field Equations on the Brane}

Further, we impose a gauge: $A_{4 \gamma} = 0$, $B_{4 \gamma} = 0$
and assume that the hypersurface $y=0$ coincides with the brane
world. Suppose that the vector normal to the brane is given by
$n_c dx^c = dy$ and $g_{ab} n^a n^b = 1$; then
\begin{equation}
ds_5^2 = q_{\mu \nu} dx^\mu dx^\nu + dy^2.
\end{equation}
Consider the Gauss like equation (\ref{g-eqs}) in a space with
nontrivial torsion
\begin{equation}\label{gc-eqs}
\dimi{4}R_{\alpha \beta} = R_{cd} e^c_\alpha e^d_\beta + K_{4
\alpha \beta} K - K_{4 \alpha \gamma} K^{4 \gamma}_{\dsps \beta} -
R^f_{\sps ced} e^c_\alpha e^d_\beta n^e n_f,
\end{equation}
where $K_{4 \alpha \beta} = -e^c_\alpha e^d_\beta \ecnabla{d}n_{c}
\equiv \es{K}_{4 \alpha \beta} + \Theta_{4 \alpha \beta}$ and $K
\equiv K_{4 \lambda}^{\dsps \lambda}$.
\newline The decomposition of curvatures into
Riemannian and non-Riemannian parts is as follows
\begin{equation}\label{decomp3}
R_{ab} \equiv \es{R}_{ab} + \widetilde{R}_{ab} = \es{R}_{ab} +
\left( \enabla{c} \Theta^c_{\sps ab} - \Theta_{acd} \Theta_b^{\sps
cd} \right), \; R \equiv \es{R} + \widetilde{R} = \es{R} -
\Theta^2.
\end{equation}
The terms $R^a_{\sps cbd} n_a e^c_\alpha e^d_\beta n^b$,
$R^c_{\sps def} n_c n^d n^e n^f$, and $R^f_{\sps abc} n_f n^a n^b
e^c_\mu$ are given by equations
\begin{subequations}
\begin{eqnarray}\label{some1}
e_\gamma^a \ecnabla{[\mu}\ecnabla{4]} n_a &=& e_\gamma^a e^c_\mu
n^d \ecnabla{[c} \ecnabla{d]} n_a + \fracmy{1}{2} e_\gamma^a
\left( \ecnabla{\mu} n^c - \ecnabla{4} e^c_\mu \right) \ecnabla{c}
n_a = \fracmy{1}{2} R^f_{\dsps adc} n_f e_\gamma^a n^d e^c_\mu,
\notag \\ &=& \fracmy{1}{2} e^a_\gamma n^c \ecnabla{c} \left(
e^\lambda_a K_{4 \lambda \mu} \right) = \fracmy{1}{2} \left(
\partial_4 K_{4 \gamma \mu} + K_{4 \lambda \mu} K_{4
\gamma}^{\dsps \lambda}\right); \label{some1a} \\ n^a
\ecnabla{[\mu}\ecnabla{4]} n_a &=& n^a e^c_\mu n^d \ecnabla{[c}
\ecnabla{d]} n_a + \fracmy{1}{2} n^a \left( \ecnabla{\mu} n^c -
\ecnabla{4} e^c_\mu \right) \ecnabla{c} n_a = \fracmy{1}{2}
R^f_{\dsps adc} n_f n^a n^d e^c_\mu, \notag \\ &=& 0; \quad
\label{some1d} \\ n_a \ecnabla{[4} \ecnabla{4]} n^a &=& n_a n^c
\ecnabla{[c} (n^d \ecnabla{d]} n^a) = \fracmy{1}{2} R^a_{\sps fcd}
n_a n^f n^c n^d, \notag \\ &=& 0. \label{some1c}
\end{eqnarray}
\end{subequations}
Following methods of \cite{br1a,br1b} we evaluate effective field
equations not exactly on the brane but at the limiting values of
$y$. In other words, we consider two sets of effective equations
denoted by the $^\pm$ sign and have taken at $y \rightarrow \pm
0$. In this case, the delta function vanishes; then using
(\ref{f-eqs1}), (\ref{gc-eqs}), (\ref{decomp3}), and
(\ref{some1c}) it can straightforwardly be checked that
\begin{eqnarray}\label{eff-eqs1}
\dimi{4}G_{(\mu \nu)}^{\pm} &\equiv& \left| \dimi{4}R_{(\mu \nu)}
- \frac{1}{2}q_{\mu \nu} \dimi{4}R - 2\Theta_{\mu \alpha \beta}
\Theta_{\nu}^{\sps \alpha \beta} \right|^{\pm} \\ &=& \left|
\fracmy{2}{3}k^2 \left[ T_{\mu \nu} + \left( T_{44} -
\fracmy{1}{4}T_c^c \right)q_{\mu \nu} \right] - C_{4 \mu 4 \nu}
\right. \notag \\ &+& K_{4 (\mu \nu)} K - K_{4 (\mu |\gamma|} K^{4
\gamma}_{\tsps \nu)} - \fracmy{1}{2}q_{\mu \nu} \left( K^2 - K_{4
\alpha \beta} K^{4 \alpha \beta} \right) \notag \\ &-& \left.
\fracmy{1}{4} \Theta_{\alpha \beta \gamma} \Theta^{\alpha \beta
\gamma} q_{\mu \nu} - \Theta_{\mu \alpha \beta} \Theta_{\nu}^{\sps
\alpha \beta} + \fracmy{1}{4} \Theta_{4 \alpha \beta} \Theta^{4
\alpha \beta} q_{\mu \nu} + 3 \Theta_{4 \mu \gamma} \Theta_{4
\nu}^{\dsps \gamma} \right|^{\pm}, \notag
\end{eqnarray}
\begin{equation}\label{weyl-riemm}
\text{where} \hspace{1cm} C_{abcd} \equiv \es{R}_{abcd} +
\fracmy{2}{3}\left( g_{a[d}\es{R}_{c]b} + g_{b[c}\es{R}_{d]a}
\right) + \fracmy{1}{6}\es{R} g_{a[c}g_{d]b} \notag
\end{equation}
is by definition the 5-dimensional Weyl tensor \cite{misc1,misc3}.
The $Z_2$-symmetry implies that
\begin{subequations}\label{z2}
\begin{eqnarray}
K_{4 (\mu \nu)}^{+} &=& - K_{4 (\mu \nu)}^{-}, \quad H_{4 \mu
\nu}^{+} = - H_{4 \mu \nu}^{-}, \quad F_{4 \mu \nu}^{+} = - F_{4
\mu \nu}^{-}; \\ \partial_4 \rho^{+} &=& - \partial_4 \rho^{-},
\quad \partial_4 \phi^{+} = - \partial_4 \phi^{-}, \quad
\partial_4 \chi^{+} = - \partial_4 \chi^{-}.
\end{eqnarray}
\end{subequations}
Hence we can drop the $^{\pm}$ sign and evaluate the equation
(\ref{eff-eqs1}) on the one side only. \newline Taking into
account (\ref{some8}), (\ref{some9}) the effective energy-momentum
tensor can be expressed as
\begin{equation}\label{eff-matter}
T_{\mu \nu} + \left( T_{44} - \fracmy{1}{4}T_c^c \right)q_{\mu
\nu} = \fracmy{3}{2} \left( -\fracmy{1}{2} \Lambda q_{\mu \nu} +
k^{-2} T_{\mu \nu}^{\Sigma} + k^2 T_{\mu \nu}^{\bot} \right),
\quad \text{where}
\end{equation}
\begin{eqnarray}
T_{\mu \nu}^{\Sigma} &\equiv& \fracmy{1}{3} \partial_\mu \phi
\partial_\nu \phi + \fracmy{5}{12} \partial_{(\mu} \phi
\partial_{\nu)} \rho + \fracmy{5}{24} \partial_\mu \rho \partial_\nu
\rho + \fracmy{e^{2\phi + \fracmy{5}{4}\rho}}{3} \partial_\mu \chi
\partial_\nu \chi + \fracmy{e^{-\phi}}{6} H_{\mu \alpha \beta}
H_{\nu}^{\sps \alpha \beta} \notag \\ && - \fracmy{q_{\mu
\nu}}{24} \left( 5 \partial_\alpha \phi \partial^\alpha \phi +
\fracmy{25}{4} \partial_\alpha \phi \partial^\alpha \rho +
\fracmy{25}{8} \partial_\alpha \rho \partial^\alpha \rho + 5
e^{2\phi + \fracmy{5}{4}\rho} \partial_\alpha \chi \partial^\alpha
\chi + \fracmy{3 e^{-\phi}}{2} H_{\alpha \beta \gamma} H^{\alpha
\beta \gamma} \right); \notag
\end{eqnarray}
\begin{eqnarray}
T_{\mu \nu}^{\bot} &\equiv& k^{-4} \left[ \fracmy{e^{-\phi}}{3}
H_{\mu \alpha 4} H_{\nu}^{\sps \alpha 4} + \fracmy{q_{\mu
\nu}}{24} \left( 3\partial_4 \phi \partial_4 \phi + \fracmy{15}{4}
\partial_4 \phi \partial_4 \rho + \fracmy{15}{8}
\partial_4 \rho \partial_4 \rho + 3 e^{2\phi + \fracmy{5}{4}\rho}
\partial_4 \chi \partial_4 \chi \right. \right. \notag \\ &&
\left. \left. - \fracmy{e^{-\phi}}{2} H_{\alpha \beta 4} H^{\alpha
\beta 4} \right) - \fracmy{1}{8} q_{\mu \nu} e^{\fracmy{5}{4}\rho}
\varrho^2 + \fracmy{1}{4} q_{\mu \nu} e^{-\fracmy{\rho}{2}} \alpha
\right]. \notag
\end{eqnarray}
Clearly, that the effective equation (\ref{eff-eqs1}) is not
closed in four dimensions due to the presence of extrinsic
quantities $\Theta_{\alpha \beta 4}$, $C_{4 \mu 4 \nu}$, and
$T_{\mu \nu}^{\bot}$. However both the equations (\ref{f-eqs1}) -
(\ref{f-eqs6}) and the assumed continuity of fields $(g_{ab},
\rho, \phi, \chi, A_{ab}, B_{ab})$ lead to Israel's like junction
conditions \cite{misc2}. These conditions gives a possibility to
decrease the number of extrinsic quantities.

The mentioned junction conditions can be expressed as follows
\begin{subequations}\label{junc-cond}
\begin{equation}
\{ g_{ab} \} = 0; \quad \{ \rho \} = 0; \quad \{ \phi \} = 0;
\quad \{ \chi \} = 0; \quad \{ B_{ab} \} = 0; \quad \{ A_{ab} \} =
0, \notag
\end{equation}
\begin{equation}
\{ K_{4 (\mu \nu)} \} = k^2 \left( S_{\mu \nu} - \fracmy{1}{3}
q_{\mu \nu} S_\lambda^\lambda \right),
\end{equation}
\begin{equation}
\{ \partial_4 \phi \} + \{ \partial_4 \rho \} =
\fracmy{k^2}{2\sqrt{3}} e^{\fracmy{1}{2}\left( \phi +
\fracmy{5}{4}\rho \right)} \theta^{\mu \alpha \nu}
\widetilde{F}_{\alpha \mu \nu},
\end{equation}
\begin{equation}
\{ \partial_4 \phi \} + \fracmy{5}{8} \{ \partial_4 \rho \} =
\fracmy{k^2}{4\sqrt{3}} e^{\fracmy{1}{2}\left( \phi +
\fracmy{5}{4}\rho \right)} \theta^{\mu \alpha \nu}
\widetilde{F}_{\alpha \mu \nu},
\end{equation}
\begin{equation}
\{ \partial_4 \chi \} = \fracmy{k^2}{2\sqrt{3}}
e^{-\fracmy{1}{2}\left( 3\phi + \fracmy{5}{4}\rho \right)}
\theta^{\alpha \mu \nu} H_{\alpha \mu \nu},
\end{equation}
\begin{equation}
\{ F_{4 \mu \nu} \} - \chi \{ H_{4 \mu \nu} \} = 2\sqrt{3} \: k^2
\: e^{-\fracmy{1}{2}\left( \phi + \fracmy{5}{4}\rho \right)}
\left[ \theta_{\mu \nu}^{\dsps \gamma} \: \partial_\gamma
\left(\phi + \fracmy{5}{4}\rho \right) + 2 \: \sigma_{\mu
\nu}\right],
\end{equation}
\begin{eqnarray}
\chi \{ F_{4 \mu \nu} \} - \left( e^{-2\phi - \fracmy{5}{4}\rho} +
\chi^2 \right) \{ H_{4 \mu \nu} \} &=& \sqrt{3} \: k^2 \:
e^{-\fracmy{1}{2}\left( \phi + \fracmy{5}{4}\rho \right)} \left[
\chi \: \sigma_{\mu \nu} \right. \notag \\ && + \left. \theta_{\mu
\nu}^{\dsps \gamma} \left( \partial_\gamma \chi + \fracmy{1}{2}
\chi \partial_\gamma \phi + \fracmy{5}{8} \chi
\partial_\gamma \rho \right) \right], \quad
\end{eqnarray}
\end{subequations}
where $\{\mathcal{V}\} \equiv \lim_{y \rightarrow +0} \mathcal{V}
- \lim_{y \rightarrow -0} \mathcal{V} \equiv \mathcal{V}^{+} -
\mathcal{V}^{-}$ denotes the discontinuity between two sides of
the brane. The $Z_2$-symmetry (\ref{z2}) implies that we can once
again drop the $^{+}$ sign and evaluate quantities on the brane by
taking only the limit $y \rightarrow +0$. The junction conditions
(\ref{junc-cond}) become
\begin{subequations}\label{junc-conds}
\begin{equation}\label{junc-conds1}
K_{4 (\mu \nu)} = \fracmy{k^2}{2} \left( S_{\mu \nu} -
\fracmy{1}{3} q_{\mu \nu} S_\lambda^\lambda \right),
\end{equation}
\begin{equation}
\partial_4 \rho = \fracmy{k^2}{3\sqrt{3}}
e^{\fracmy{1}{2}\left( \phi + \fracmy{5}{4}\rho \right)}
\theta^{\mu \alpha \nu} \widetilde{F}_{\alpha \mu \nu},
\end{equation}
\begin{equation}
\partial_4 \phi = \fracmy{k^2}{12\sqrt{3}}
e^{\fracmy{1}{2}\left( \phi + \fracmy{5}{4}\rho \right)}
\theta^{\alpha \mu \nu} \widetilde{F}_{\alpha \mu \nu},
\end{equation}
\begin{equation}
\partial_4 \chi = \fracmy{k^2}{4\sqrt{3}}
e^{-\fracmy{1}{2}\left( 3\phi + \fracmy{5}{4}\rho \right)}
\theta^{\alpha \mu \nu} H_{\alpha \mu \nu},
\end{equation}
\begin{equation}
H_{4 \mu \nu} = \fracmy{\sqrt{3}}{2} \: k^2 \:
e^{\fracmy{1}{2}\left( 3\phi + \fracmy{5}{4}\rho \right)} \left(
3\chi \: \sigma_{\mu \nu} - \theta_{\mu \nu}^{\dsps \gamma}
\Pi_\gamma \right),
\end{equation}
\begin{eqnarray}
F_{4 \mu \nu} &=& \fracmy{\sqrt{3}}{2} \: k^2 \:
e^{\fracmy{1}{2}\left( 3\phi + \fracmy{5}{4}\rho \right)} \left[
\theta_{\mu \nu}^{\dsps \gamma} \left( 2 e^{-\left(
2\phi + \fracmy{5}{4}\rho \right)} \widehat{\Pi}_\gamma - \chi
\Pi_\gamma \right) + \sigma_{\mu \nu} \left( 4 e^{-\left(
2\phi + \fracmy{5}{4}\rho \right)} + 3 \chi^2 \right) \right],
\hspace{1.3cm}
\end{eqnarray}
\end{subequations}
\begin{equation}
\text{where} \hspace{1cm} \Pi_\gamma \equiv \left( \partial_\gamma
\chi - \fracmy{3}{2} \chi \partial_\gamma \phi - \fracmy{15}{8}
\chi \partial_\gamma \rho \right); \quad \widehat{\Pi}_\gamma
\equiv \partial_\gamma \left(\phi + \fracmy{5}{4}\rho \right).
\notag
\end{equation}
In other words, these conditions mean that quantities $(K_{4 (\mu
\nu)}, H_{4 \mu \nu}, F_{4 \mu \nu},
\partial_4 \rho, \partial_4 \phi, \partial_4 \chi)$ depend on the
brane matter content. In addition, these conditions make possible
to express $\Theta_{4 \mu \nu}$ and $T_{\mu \nu}^{\bot}$ as
follows
\begin{equation}\label{rule2}
K_{4 [\mu \nu]} = \Theta_{4 \mu \nu} = \fracmy{1}{2} k^2 \left(
\theta_{\mu \nu}^{\dsps \gamma} \: \widehat{\Pi}_\gamma + 2 \:
\sigma_{\mu \nu} \right),
\end{equation}
\begin{eqnarray}
T_{\mu \nu}^{\bot} &=& e^{\phi + \fracmy{5}{4}\rho} \left[
\fracmy{e^{\phi}}{4} \left( \theta_{\mu \lambda}^{\dsps \alpha}
\theta_{\nu}^{\sps \lambda \beta} \Pi_\alpha \Pi_\beta - 6 \chi
\theta_{(\mu}^{\sps \lambda \alpha} \sigma_{\nu) \lambda}
\Pi_\alpha + 9 \chi^2 \sigma_{\mu \lambda} \sigma_{\nu}^{\sps
\lambda} \right) \right. \notag \\ && - \fracmy{e^{\phi}}{64}
q_{\mu \nu} \left( \theta_{\alpha \beta}^{\dsps \gamma}
\theta^{\alpha \beta \lambda} \Pi_\gamma \Pi_\lambda - 6 \chi
\sigma_{\alpha \beta} \theta^{\alpha \beta \gamma} \Pi_\gamma + 9
\chi^2 \sigma_{\alpha \beta} \sigma^{\alpha \beta} \right)  +
\fracmy{1}{576}q_{\mu \nu} \left( \theta^{\gamma \alpha \beta}
\widetilde{F}_{\gamma \alpha \beta} \right)^2 \notag \\ && \left.
+ \fracmy{e^{-2\phi - \fracmy{5}{4}\rho}}{384} q_{\mu \nu} \left(
\theta^{\gamma \alpha \beta} H_{\gamma \alpha \beta} \right)^2
\right] - \fracmy{1}{8} q_{\mu \nu} e^{\fracmy{5}{4}\rho}
\widetilde{\varrho}^2 + \fracmy{1}{4} q_{\mu \nu}
e^{-\fracmy{\rho}{2}} \widetilde{\alpha},
\end{eqnarray}
where $\widetilde{\alpha} \equiv \alpha k^{-4}$ and
$\widetilde{\varrho} \equiv \varrho k^{-2}$.
\newline Further, we conclude that the equation (\ref{eff-eqs1})
contains extrinsic quantities only in $C_{4 \mu 4 \nu}$ term. It
is a projection of the bulk Weyl tensor, which vanishes whenever
the bulk is exactly $AdS_5$.

At this stage we apply the relation (\ref{rule1}) in order to
consider one dynamically equivalent to (\ref{eff-eqs1}) equation
on the Riemannian space instead of EC space.
\begin{eqnarray}\label{eff-eqs2}
\dimi{4}\es{G}_{\mu \nu} &\equiv& \dimi{4}G_{(\mu \nu)} -
\fracmy{1}{2} q_{\mu \nu} \Theta_{\alpha \beta \gamma}
\Theta^{\alpha \beta \gamma} + 3\Theta_{\mu \alpha \beta}
\Theta_\nu^{\sps \alpha \beta} \\ &=& \fracmy{2}{3}k^2 \left[
T_{\mu \nu} + \left( T_{44} - \fracmy{1}{4}T_c^c \right)q_{\mu
\nu} \right] - C_{4 \mu 4 \nu} \notag \\ &+& \es{K}_{4 \mu \nu}
\es{K} - \es{K}_{4 \mu \gamma} \esi{K^{4 \gamma}_{\tsps \nu}} -
\fracmy{1}{2}q_{\mu \nu} \left( \es{K^2} - \es{K}_{4 \alpha \beta}
\esi{K^{4 \alpha \beta}} \right) \notag
\\ &-& \fracmy{3}{4} \Theta_{\alpha \beta \gamma} \Theta^{\alpha
\beta \gamma} q_{\mu \nu} + 2\Theta_{\mu \alpha \beta}
\Theta_{\nu}^{\sps \alpha \beta} - \fracmy{1}{4} \Theta_{4 \alpha
\beta} \Theta^{4 \alpha \beta} q_{\mu \nu} + 4 \Theta_{4 \mu
\gamma} \Theta_{4 \nu}^{\dsps \gamma}. \notag
\end{eqnarray}
This equation can be compared to the restricted result
\cite{br1a,br1b}, where only gravity exists in the bulk. Recall
that the brane energy-momentum tensor $\tau_{\mu \nu}$ is
decomposed into a sum of $\es{\tau}_{\mu \nu}$ and
$\widetilde{\tau}_{\mu \nu}$ parts given by (\ref{emt-decomp}),
where $\widetilde{\tau}_{\mu \nu}$ describes interactions with
bulk fields localized on the brane. Taking account of
(\ref{junc-conds}), (\ref{rule1}), (\ref{rule2}), and
(\ref{eff-matter}) the equation (\ref{eff-eqs2}) becomes
\begin{eqnarray}\label{5d-brane-eqs1}
\dimi{4}\es{G}_{\mu \nu} = - \widehat{\Lambda} q_{\mu \nu} + 8 \pi
G_N \es{\tau}_{\mu \nu} + \: \widehat{T}^{\Sigma}_{\mu \nu} + k^4
\left( \es{\pi}_{\mu \nu} + \: \widehat{T}^{\bot}_{\mu \nu}
\right) - E^{g}_{\mu \nu},
\end{eqnarray}
\begin{eqnarray}
\text{where} \quad \widehat{\Lambda} &\equiv& \fracmy{1}{2} k^2
\left( \Lambda + \fracmy{1}{6} \lambda k^2 \right), \notag \\ G_N
&\equiv& \fracmy{\lambda k^4}{48 \pi}, \notag \\ \es{\pi}_{\mu
\nu} &\equiv& -\fracmy{1}{4} \es{\tau}_{\mu \alpha} \es{\tau}_{\nu
\beta} q^{\alpha \beta} + \fracmy{1}{12} \es{\tau} \es{\tau}_{\mu
\nu} + \fracmy{1}{8} q_{\mu \nu} \es{\tau}_{\alpha \gamma}
\es{\tau}_{\beta \lambda} q^{\alpha \beta} q^{\gamma \lambda} -
\fracmy{1}{24} q_{\mu \nu} \es{\tau} \es{\tau}, \hspace{1cm}
\notag \\ E^{g}_{\mu \nu} &\equiv& C_{4 \mu 4 \nu}, \notag
\end{eqnarray}
\begin{eqnarray}
\text{and} \quad \widehat{T}^{\Sigma}_{\mu \nu} &\equiv&
T^{\Sigma}_{\mu \nu} + \fracmy{1}{6} e^{\phi + \fracmy{5}{4}\rho}
\widetilde{F}_{\mu \alpha \beta} \widetilde{F}_{\nu}^{\sps \alpha
\beta} - \fracmy{1}{16}q_{\mu \nu} \: e^{\phi + \fracmy{5}{4}\rho}
\widetilde{F}_{\alpha \beta \gamma} \widetilde{F}^{\alpha \beta
\gamma} + \fracmy{\lambda k^4 }{6} \widetilde{\tau}_{\mu \nu},
\notag \\ \widehat{T}^{\bot}_{\mu \nu} &\equiv& T^{\bot}_{\mu \nu}
+ \left( \theta_{\mu \lambda}^{\dsps \alpha} \theta_{\nu}^{\sps
\lambda \beta} \widehat{\Pi}_{\alpha} \widehat{\Pi}_{\beta} + 4
\theta_{(\mu}^{\sps \lambda \gamma} \sigma_{\nu) \lambda}
\widehat{\Pi}_{\gamma} + 4 \sigma_{\mu \gamma} \sigma_{\nu}^{\sps
\gamma} \right) \notag \\ && - \fracmy{q_{\mu \nu}}{16} \left(
\theta_{\alpha \beta}^{\dsps \gamma} \theta^{\alpha \beta \lambda}
\widehat{\Pi}_{\gamma} \widehat{\Pi}_{\lambda} + 4 \theta_{\alpha
\beta}^{\dsps \lambda} \sigma^{\alpha \beta}
\widehat{\Pi}_{\lambda} + 4 \sigma_{\alpha \beta} \sigma^{\alpha
\beta} \right) \notag \\ && - \fracmy{1}{2} \es{\tau}_{\lambda
(\mu} \widetilde{\tau}_{\nu)}^{\sps \lambda} + \fracmy{1}{12}
\left( \es{\tau}_{\mu \nu} \widetilde{\tau} +
\widetilde{\tau}_{\mu \nu} \es{\tau} \right) + \fracmy{1}{4}
q_{\mu \nu} \es{\tau}_{\alpha \beta} \widetilde{\tau}^{\alpha
\beta} - \fracmy{1}{12} q_{\mu \nu} \es{\tau} \widetilde{\tau}
\notag \\ && - \fracmy{1}{4} \widetilde{\tau}_{\mu \lambda}
\widetilde{\tau}_{\nu}^{\sps \lambda} + \fracmy{1}{12}
\widetilde{\tau} \widetilde{\tau}_{\mu \nu} + \fracmy{1}{8} q_{\mu
\nu} \widetilde{\tau}_{\alpha \beta} \widetilde{\tau}^{\alpha
\beta} - \fracmy{1}{24} q_{\mu \nu} \widetilde{\tau}^2. \notag
\end{eqnarray}
The main difference of (\ref{5d-brane-eqs1}) from Einstein gravity
resides in the presence of terms $\es{\pi}_{\mu \nu}$,
$\widehat{T}^{\Sigma}_{\mu \nu}$, $\widehat{T}^{\bot}_{\mu \nu}$,
and $E^{g}_{\mu \nu}$ which can not be obtained by a Lagrangian
description. It is seen that in the limit whenever $\lambda$ is of
high energy scale the terms $\es{\pi}_{\mu \nu}$ and
$\widehat{T}^{\bot}_{\mu \nu}$ can safely be neglected.

Now let us examine the behaviour of the brane energy-momentum
tensor $S_{\mu \nu}$ [see (\ref{f-eqs1})]. Consider the Codacci
like equation (\ref{c-eqs}) in a space with nontrivial torsion
\begin{equation}\label{cd-eqs}
R_{ab} \: n^a e^b_\mu = \dimi{4} \ecnabla{\mu} K - \dimi{4}
\ecnabla{\lambda} K^{\sps \lambda}_{4 \sps \mu} - 2\Theta_{\lambda
\mu \gamma} K_4^{\sps \gamma \lambda} + R^a_{\sps bcd} \: n_a n^b
n^c e^d_\mu.
\end{equation}
Using (\ref{some1d}), (\ref{decomp3}) decompose (\ref{cd-eqs}) and
$G_{(ab)}$ as follows
\begin{eqnarray}
\es{R}_{ab} n^a e^b_\mu &=& \dimi{4} \ecnabla{\mu} \es{K} -
\dimi{4} \ecnabla{\lambda} \esi{K^{\sps \lambda}_{4 \sps \mu}} +
\dimi{4} \ecnabla{\lambda} \Theta^{\lambda}_{\sps 4 \mu} - e^b_\mu
\enabla{c} \Theta^c_{\sps 4 b} - \Theta_{\mu \alpha \beta}
\Theta_{4}^{\sps \alpha \beta} \notag \\ &=& \dimi{4} \enabla{\mu}
\es{K} - \dimi{4} \enabla{\lambda} \esi{K^{\sps \lambda}_{4 \sps
\mu}}, \notag \\ G_{(ab)} n^a e^b_\mu &=& \left( \es{G}_{ab} - 3
\Theta_{a cd} \Theta_{b}^{\sps cd} \right) n^a e^b_\mu =
\es{R}_{ab} n^a e^b_\mu - 3 \Theta_{a cd} \Theta_{b}^{\sps cd} n^a
e^b_\mu. \notag
\end{eqnarray}
These equations together with (\ref{rule1}), (\ref{f-eqs1}), and
(\ref{junc-conds}) lead to the relation
\begin{eqnarray}
\dimi{4} \enabla{\lambda} S^{\lambda}_{\sps \mu} &=& -2 k^{-2}
\left( P_{ab} + \fracmy{1}{4} e^{\phi + \fracmy{5}{4}\rho}
\widetilde{F}_{a cd} \widetilde{F}_{b}^{\sps cd} \right) n^a
e^b_\mu \notag \\ &=& e^{\fracmy{1}{2} \left(\phi +
\fracmy{5}{4}\rho \right)} \left[ \fracmy{1}{12 \sqrt{3}}
\theta^{\alpha \beta \gamma} \widetilde{F}_{\alpha \beta \gamma}
\: \partial_\mu \left( \fracmy{5}{4} \rho - \phi \right) -
\fracmy{1}{4\sqrt{3}} \sigma^{\alpha \beta \gamma} H_{\alpha \beta
\gamma} \: \partial_\mu \chi \right. \notag \\ && - \left.
\fracmy{\sqrt{3}}{4} \left( 3 \chi \sigma_{\alpha \beta} -
\theta_{\alpha \beta}^{\dsps \gamma} \Pi_\gamma \right)
H_{\mu}^{\sps \alpha \beta} - \fracmy{\sqrt{3}}{2} \left(
\theta_{\alpha \beta}^{\dsps \gamma} \widehat{\Pi}_\gamma + 2
\sigma_{\alpha \beta} \right) \widetilde{F}_{\mu}^{\sps \alpha
\beta} \right] \notag
\\ &\equiv& J_{\mu}.
\end{eqnarray}
This relation indicates that the brane energy-momentum tensor
$S_{\mu \nu}$ is non-conserved. The non-conservation reflects an
exchange of energy-momentum between the brane and bulk fields
localized on the brane (see \cite{br4} for related work). If one
imposes $J_{\mu} = 0$; then there is no such exchange and the
brane vacuum state remains stable. However the effective brane
energy-momentum tensor given by the right part of
(\ref{5d-brane-eqs1}) is conserved and the effective vacuum
remains stable. Let us remark that throughout the literature, the
true brane energy-momentum tensor in presence of bulk fields is
understood ambiguously.

By the same arguments, the equations (\ref{f-eqs2}) -
(\ref{f-eqs6}) must also be taken at the limiting values of $y$.
In this case, the situation is simpler. Taking into account
(\ref{s6}) one see that
\begin{eqnarray}
\dimi{5}\enabla{c} \hspace{-0.2cm} \dimi{5}\uenabla{c} \phi &=&
\dimi{4}\enabla{\lambda} \hspace{-0.2cm} \dimi{4}\uenabla{\lambda}
\phi - K \partial_4 \phi + \partial_4 \partial_4 \phi, \notag \\
\dimi{5}\enabla{c} F^{c \mu \nu} &=& \dimi{4}\enabla{\lambda}
F^{\lambda \mu \nu} - K F^{4 \mu \nu}  + \partial_4 F^{4 \mu \nu};
\notag
\end{eqnarray}
then using (\ref{junc-conds}), (\ref{f-eqs7}) and (\ref{some9})
equations (\ref{f-eqs2}) - (\ref{f-eqs6}) become
\begin{subequations}\label{5d-brane-eqs}
\begin{eqnarray}\label{5d-brane-eqs2}
&& \dimi{4}\enabla{\lambda} \hspace{-0.2cm}
\dimi{4}\uenabla{\lambda} \left( \rho + \phi \right) -
e^{\fracmy{5}{4}\rho} \left( e^{2\phi} \partial_\gamma \chi
\partial^\gamma \chi + \fracmy{e^\phi}{6} \widetilde{F}_{\alpha \beta \gamma}
\widetilde{F}^{\alpha \beta \gamma} \right) +
e^{\fracmy{5}{4}\rho} \varrho^2 - \fracmy{4}{5}
e^{-\fracmy{1}{2}\rho} \alpha + \left( E_{\phi} + E_{\rho} \right)
\notag \\ && = k^4 \left[ \fracmy{e^{-\phi}}{16\sqrt{3}} \left(
\theta^{\alpha \mu \nu} H_{\alpha \mu \nu} \right)^2 +
\fracmy{3}{2} \left( \theta_{\alpha \beta}^{\dsps \gamma}
\theta^{\alpha \beta \lambda} \widehat{\Pi}_\gamma
\widehat{\Pi}_\lambda + 4 \theta^{\alpha \beta \gamma}
\widehat{\Pi}_\gamma \: \sigma_{\alpha \beta} + 4 \sigma_{\alpha
\beta} \sigma^{\alpha \beta}\right) \right. \notag
\\ && + \left. \fracmy{e^{\fracmy{1}{2}\left(\phi + \fracmy{5}{4}\rho
\right)}}{24\sqrt{3}} S_{\lambda}^{\lambda} \: \theta^{\alpha \mu
\nu} \widetilde{F}_{\alpha \mu \nu} \right],
\end{eqnarray}
\begin{eqnarray}\label{5d-brane-eqs3}
&& \dimi{4}\enabla{\gamma} \hspace{-0.2cm} \dimi{4}
\uenabla{\gamma} \left( \phi + \fracmy{5}{8} \rho \right) -
e^{\fracmy{5}{4}\rho} \left( e^{2\phi} \partial_\gamma \chi
\partial^\gamma \chi + \fracmy{e^{\phi}}{12} \widetilde{F}_{\alpha
\beta \gamma} \widetilde{F}^{\alpha \beta \gamma} \right) +
\fracmy{e^{-\phi}}{12} H_{\alpha \beta \gamma} H^{\alpha \beta
\gamma} \notag \\ && + \left( E_{\phi} + \fracmy{5}{8} E_{\rho}
\right) = k^4 \left[ \fracmy{3}{4} e^{2\phi + \fracmy{5}{4}\rho}
\left( 6 \theta^{\alpha \beta \gamma} \Pi_\gamma \chi \:
\sigma_{\alpha \beta} - \theta_{\alpha \beta}^{\dsps \gamma}
\theta^{\alpha \beta \lambda} \Pi_\gamma \Pi_\lambda - 9 \chi^2
\sigma_{\alpha \beta} \sigma^{\alpha \beta} \right) \right. \notag
\\ && + \fracmy{e^{\fracmy{1}{2}\left(\phi + \fracmy{5}{4}\rho
\right)}}{48\sqrt{3}} S_{\lambda}^{\lambda} \: \theta^{\alpha \mu
\nu} \widetilde{F}_{\alpha \mu \nu} + \fracmy{3}{4} \left(
\theta_{\alpha \beta}^{\dsps \gamma} \theta^{\alpha \beta \lambda}
\widehat{\Pi}_\gamma \widehat{\Pi}_\lambda + 4 \theta^{\alpha
\beta \gamma} \widehat{\Pi}_\gamma \: \sigma_{\alpha \beta} + 4
\sigma_{\alpha \beta} \sigma^{\alpha \beta} \right) \notag \\ &&
\left. + \fracmy{e^{-\phi}}{16\sqrt{3}} \left( \theta^{\alpha \mu
\nu} H_{\alpha \mu \nu} \right)^2 \right],
\end{eqnarray}
\begin{eqnarray}\label{5d-brane-eqs4}
&& \dimi{4}\enabla{\gamma} \hspace{-0.2cm} \dimi{4}
\uenabla{\gamma} \chi + 2 \partial^\gamma \left( \phi +
\fracmy{5}{8}\rho \right) \partial_\gamma \chi +
\fracmy{e^{-\phi}}{6} \widetilde{F}_{\alpha \beta \gamma}
H^{\alpha \beta \gamma} + E_\chi \notag \\ && = k^4 \left[
\fracmy{ e^{-\phi}}{32\sqrt{3}} \theta^{\lambda \mu \nu}
H_{\lambda \mu \nu} \theta^{\gamma \alpha \beta}
\widetilde{F}_{\gamma \alpha \beta} - \fracmy{e^{-\fracmy{1}{2}
\left(3\phi + \fracmy{5}{4}\rho \right)}}{24\sqrt{3}}
S_{\lambda}^{\lambda} \: \theta^{\alpha \mu \nu} H_{\alpha \mu
\nu} \right. \notag \\ && \left. + \fracmy{3}{4} \left(
\theta_{\alpha \beta}^{\dsps \gamma} \theta^{\alpha \beta \lambda}
\widehat{\Pi}_\gamma \Pi_\lambda - 3 \chi \theta^{\alpha \beta
\gamma} \widehat{\Pi}_\gamma \sigma_{\alpha \beta} + 2
\theta^{\alpha \beta \gamma} \Pi_\gamma \: \sigma_{\alpha \beta} -
6 \chi \sigma_{\alpha \beta} \sigma^{\alpha \beta} \right)
\right],
\end{eqnarray}
\begin{eqnarray}\label{5d-brane-eqs5}
&& \dimi{4}\enabla{\gamma} \widetilde{F}^{\gamma \mu \nu} +
\partial_\gamma \left( \phi + \fracmy{5}{4} \rho \right)
\widetilde{F}^{\gamma \mu \nu} + E^{\mu \nu}_{\widetilde{F}} = k^4
\left[ \fracmy{\sqrt{3}}{4} e^{\fracmy{1}{2}\left( \phi -
\fracmy{5}{4}\rho \right)} \eta^{\mu \nu \alpha \beta} \left(
3 \chi \sigma_{\alpha \beta} - \theta_{\alpha \beta}^{\dsps \gamma} \Pi_\gamma
\right) \widetilde{\varrho} \right. \notag
\\ && \left. - \fracmy{1}{3} \left( \theta^{\mu \nu \gamma}
\widehat{\Pi}_\gamma + 2 \sigma^{\mu \nu} \right) \left(
\fracmy{\sqrt{3}}{2} e^{-\fracmy{1}{2}\left( \phi +
\fracmy{5}{4}\rho \right)} S^\lambda_\lambda + \theta^{\beta
\alpha \gamma} \widetilde{F}_{\alpha \beta \gamma} \right)
\right],
\end{eqnarray}
\begin{eqnarray}\label{5d-brane-eqs6}
&& \dimi{4} \enabla{\gamma} \left( e^{-\phi} H^{\gamma \mu \nu} -
\chi \: e^{\phi + \fracmy{5}{4}\rho} \widetilde{F}^{\gamma \mu
\nu} \right) + e^{-\phi} E^{\mu \nu}_{H} - \chi e^{\phi +
\fracmy{5}{4}\rho} E^{\mu \nu}_{\widetilde{F}} \\ && = k^4
e^{\fracmy{1}{2}\left( \phi + \fracmy{5}{4} \rho \right)} \left[
\fracmy{1}{12} \left( \sqrt{3} \: S^\lambda_\lambda -
\fracmy{1}{2} e^{\fracmy{1}{2}\left( \phi + \fracmy{5}{4} \rho
\right)} \theta^{\alpha \beta \gamma} \widetilde{F}_{\alpha \beta
\gamma} \right) \left( \theta^{\mu \nu \gamma} \Pi_\gamma - 3\chi
\sigma^{\mu \nu} \right) \right. \notag \\ && + \left(
\fracmy{1}{4} e^{-\fracmy{1}{2}\left( 3\phi + \fracmy{5}{4} \rho
\right)} \theta^{\alpha \beta \gamma} H_{\alpha \beta \gamma} -
\fracmy{\chi}{3} \: e^{\fracmy{1}{2}\left( \phi + \fracmy{5}{4}
\rho \right)} \: \theta^{\alpha \beta \gamma}
\widetilde{F}_{\alpha \beta \gamma} + \fracmy{\chi}{2 \sqrt{3}}
S^\lambda_\lambda \right) \left( \theta^{\mu \nu \gamma}
\widehat{\Pi}_\gamma + 2 \sigma^{\mu \nu} \right) \notag \\ &&
\left. + \fracmy{\sqrt{3}}{4} \widetilde{\varrho} \: \eta^{\mu
\nu\alpha \beta} e^{\phi} \left[ \theta_{\alpha \beta}^{\dsps
\gamma} \left( 2 e^{-\left( 2\phi + \fracmy{5}{4}\rho \right)}
\widehat{\Pi}_\gamma - \chi \Pi_\gamma \right) + \sigma_{\alpha
\beta} \left( 4 e^{-\left( 2\phi + \fracmy{5}{4}\rho \right)} + 3
\chi^2 \right) \right] \right] \notag,
\end{eqnarray}
\end{subequations}
\begin{equation}
\text{where} \quad E_{\phi} \equiv \partial_4 \partial_4 \phi,
\quad E_{\rho} \equiv
\partial_4 \partial_4 \rho, \quad E_{\chi} \equiv \partial_4 \partial_4
\chi, \quad E_{\mu \nu}^{H} \equiv \partial_4 H_{4 \mu \nu}, \quad
E_{\mu \nu}^{\widetilde{F}} \equiv \partial_4 \widetilde{F}_{4 \mu
\nu}, \notag
\end{equation}
and $\eta_{\alpha \beta \gamma \lambda}$ is the brane volume form
[see (\ref{some9})]. The equations (\ref{5d-brane-eqs1}),
(\ref{5d-brane-eqs2}) - (\ref{5d-brane-eqs6}), and (\ref{f-eqs8})
are the effective equations on the brane.

For the same reason, we can apply the relation (\ref{rule1}) in
order to consider these equations on the Riemannian space instead
of EC space. This gives a set of dynamically equivalent equations
of motion. At the same time notice that (\ref{5d-brane-eqs1}),
(\ref{5d-brane-eqs2}) - (\ref{5d-brane-eqs6}) have completely
described in Riemannian terms. The equation (\ref{f-eqs8}) can
readily be expressed as follows
\begin{equation}\label{some13}
\fracmy{\partial L}{\partial \psi_A} - \dimi{4} \enabla{\lambda}
\fracmy{\partial L}{\partial \dimi{4} \ecnabla{\lambda} \psi_A} =
\fracmy{\sqrt{3}}{2} e^{\fracmy{1}{2}\left(\phi +
\fracmy{5}{4}\rho\right)} \left( \partial_{[\alpha}A_{\beta
\lambda]} - \chi \: \partial_{[\alpha}B_{\beta \lambda]} \right)
\Omega^{A}_{\sps B}|^{\alpha \beta} \fracmy{\partial L}{\partial
\dimi{4} \ecnabla{\lambda} \psi_B}.
\end{equation}

\section{Conclusions}

In this paper we have derived effective field equations on the
3-brane motivated by the massless bosonic sector of the type $IIB$
string. This setup supposes that fields of this sector can
penetrate in the extra dimension. The brane worlds motivated by
string/M-theory have become an widely investigated topic in the
literature. However in our case, we have introduced the
Einstein-Cartan space in order to induce interactions between the
brane and bulk fields. By the developed embedding procedure these
interactions are understood in a purely geometric way as the
induced brane contortion. Hence this approach allows us to avoid
additional suppositions about the form of interactions. Finally,
we have presented the dynamically equivalent effective equations
expressed completely in Riemannian terms.

It has been expected that the effective theory is not closed in
4-dimensions. The considered approach does not give a possibility
to establish effective equations in terms of quantities defined on
the brane only. In order to obtain one complete set we should
derive the equations of motion for extrinsic quantities
$\{E_{\phi}, E_{\rho}, E_{\chi}, E^{H}_{\mu \nu},
E^{\widetilde{F}}_{\mu \nu}, E^{g}_{\mu \nu} \}$ in addition to
(\ref{5d-brane-eqs1}), (\ref{5d-brane-eqs2}) -
(\ref{5d-brane-eqs6}), and (\ref{some13}). However the established
effective equations (as well as the bulk equations) appears very
complex to be solved. Hence it seems reasonable to impose some
special symmetries on the bulk fields. The imposed symmetries can
eliminate extrinsic quantities or make them negligible. On the
other hand, in the papers \cite{br1a}, where only gravity exists
in the bulk the evolution of extrinsic quantities it was found and
it was shown that these contributions are subtle. This result
makes us to believe that one similar conclusion can hold in our
case. The situation with the non-conservation of the brane
energy-momentum tensor $S_{\mu \nu}$ is also curious. We noticed
that the effective brane energy-momentum tensor given by the right
part of (\ref{5d-brane-eqs1}) conserves and the effective vacuum
remains stable. Nevertheless if $S_{\mu \nu}$ is understood as the
true brane energy-momentum tensor; then additional suppositions
about the asymptotical structure of brane are necessary (see
\cite{br1a,br4,br5} for related work). The examination of all
these problems we leave for a forthcoming paper.

We stress that in the framework of brane worlds exist articles,
where the authors have also focused on the massless bosonic sector
of the type $IIB$ theory. We only mention the paper \cite{br7},
where the brane matter sector consists of a Born-Infeld action
constructed from the induced metric and projected antisymmetric
bulk fields. One motivation to consider such action is the
magnetogenesis in early universe. Let us also remark articles,
where the scenario assumes the torsion. First notice the papers
\cite{br2}, where the torsion has given by the Kalb-Ramond field
that coexists with gravity in a 6-dimensional bulk. The
4-dimensional effective theory is constructed. However this
approach is based on the Kaluza-Klein like reduction instead of
the covariant embedding formalism. Apparently, the Gauss and
Codacci like formulas obtained in our paper can easily be adapted
for the 6-dimensional case, and some new ideas can be suggested.
Finally notice the papers \cite{br3}, where the presence of bulk
fermions induces the torsion. The authors have shown that the
induced contact interaction on the brane is suppressed only by the
square of the fundamental scale, which could be of the observed
order. The authors as well do not apply the embedding formalism.

Anyway the arguments in this paper extend the results in the
literature on the brane worlds motivated by the string/M-theory.

\begin{acknowledgments}
The author is grateful to V. Franke, V. Lyakhovsky, and S. Vacaru
for collaboration and helpful discussions. The author thanks also
E. Prokhvatilov and S. Paston for support.
\end{acknowledgments}

\appendix
\section{Embedding of a Brane}\label{app1}

The geometry of embedded branes has been widely investigated in
the literature. In this appendix we construct the embedding
formalism for spaces with nontrivial torsion (see \cite{tors5} for
related work). Let $x^a(z,y)$ be the coordinates of the bulk,
$z^\lambda$ the coordinates of the brane, and $y^i$  the
coordinates of a space normal to the brane\footnote{Where the
Latin indices $a,b = 0,\ldots,D-1; \: i,j = D_\Sigma,\ldots,D-1$;
and the Greek indices $\mu,\nu = 0,\ldots,D_\Sigma-1$.}. First we
define the metric on the brane ($\Sigma$)
\begin{equation}\label{brane-metric}
q_{\mu \nu} = g_{ab}\, e_\mu^a e_\nu^b,
\end{equation}
where $e_\mu^a = \fracmy{\partial x^a}{\partial z^\mu} \equiv
\partial_\mu x^a$ is a frame associated with the intrinsic
coordinates. Properties of a non-intrinsic character encode the
unit vectors $n^a_i = \fracmy{\partial x^a}{\partial y^i} \equiv
\partial_i x^a$ normal to the brane such that $g_{ab} n^a_i
e_\mu^b = 0$. Now we define the metric of space normal to the
brane ($\bot$)
\begin{equation}\label{normal-metric}
q_{ij} = g_{ab}\, n_i^a n_j^b.
\end{equation}
The decomposition of a bulk vector is as follows
\begin{eqnarray}\label{vect-decomp}
\mathcal{V}^a = n_k^a n_b^k \mathcal{V}^b &+& e_\mu^a e_b^\mu
\mathcal{V}^b \equiv \mathcal{V}^a_\bot + \mathcal{V}^a_\Sigma
\equiv n_k^a v^k + e_\mu^a v^\mu; \\ \text{then} \quad \quad
e_\mu^a e_b^\mu = \delta_b^a - n_k^a n_b^k &\equiv& h_{\sps b}^a,
\quad \text{where} \quad e^\mu_a \equiv g_{ab}\: e_\nu^b\: q^{\mu
\nu}; \: n^i_a \equiv g_{ab}\: n_j^b\: q^{ij}, \notag \\
\text{and} \qquad h_{ab}\: e_\mu^a e_\nu^b &=& q_{\mu \nu}; \quad
h_b^{\sps a} h_c^{\sps b } = h_c^{\sps a}; \quad h_{ab} = g_{ab} -
n_{ak} n^k_b. \notag
\end{eqnarray}
These considerations mean that $h_{ab}$ is the induced metric on
the brane \cite{misc1}. Let us perform the parallel transport of
tangent vector to the brane world volume $V_a^\Sigma$ by means
$z^\gamma \rightarrow z^\gamma + dz^\gamma$ taking into account
the bulk connection $\Gamma^a_{\sps bc}$
\begin{equation}\label{s1}
V^\Sigma_{|| a}(z^\gamma + dz^\gamma) = V^\Sigma_a(z^\gamma) +
\Gamma^c_{\sps ab} V^\Sigma_c(z^\gamma) dx^b =
V^\Sigma_a(z^\gamma) + e^b_\lambda(z^\gamma) \Gamma^c_{\sps ab}
V^\Sigma_c(z^\gamma) dz^\lambda.
\end{equation}
In the basis $(n^a_i; e^a_\mu)$ the brane projection of (\ref{s1})
at $\mathcal{O}(dz^2)$ becomes
\begin{eqnarray}\label{s2}
\text{left part}&:& \quad v_{|| \mu}(z^\gamma + dz^\gamma)
e^\mu_a(z^\gamma + dz^\gamma) \simeq v_{|| \mu}(z^\gamma +
dz^\gamma) e^\mu_a(z^\gamma) + v_\mu(z^\gamma)
\partial_\gamma e^\mu_a(z^\gamma) dz^\gamma, \notag \\ \text{right
part}&:& \quad v_\mu(z^\gamma) e^\mu_a(z^\gamma) +
e^\mu_c(z^\gamma) e^b_\lambda(z^\gamma) \Gamma^c_{\sps ab}
v_\mu(z^\gamma) dz^\lambda; \notag \\ \text{finally}&:& \quad
v_{|| \mu} = v_\mu + e^a_\mu e^\nu_c e^b_\gamma \Gamma^c_{\sps ab}
v_\nu dz^\gamma - e^a_\mu v_\nu\partial_\gamma e^\nu_a dz^\gamma.
\end{eqnarray}
At the same time (\ref{s2}) defines the parallel transport rule of
brane world volume vector $v_\mu$ with respect to the brane
connection $\dimi{\Sigma} \Gamma^\lambda_{\sps \mu \nu}$ in the
form
\begin{eqnarray}\label{ind-conn}
v_{|| \mu} &=& v_\mu + \left(e^a_\mu e^b_\nu e^\lambda_c
\Gamma^c_{\sps ab} - e^c_\mu \partial_\nu e^\lambda_c \right)
v_\lambda dz^\nu \equiv v_\mu + \dimi{\Sigma} \Gamma^\lambda_{\sps
\mu \nu} v_\lambda dz^\nu, \\ \text{with} \quad \dimi{\Sigma}
\Gamma^\lambda_{\sps \mu \nu} &=& e^a_\mu e^b_\nu e^\lambda_c
\Gamma^c_{\sps ab} - e^c_\mu \partial_\nu e^\lambda_c =
e^\lambda_c \left(e^a_\mu e^b_\nu \Gamma^c_{\sps ab} +
\partial_\nu e^c_\mu\right) = g_{cd} e^d_\gamma q^{\gamma \lambda}
\ecnabla{\nu} e^c_\mu, \notag
\end{eqnarray}
where $\ecnabla{\nu} \equiv e^a_\nu \ecnabla{a}$ is the covariant
derivative along the brane world volume direction. The orthogonal
projection of (\ref{s1}) defines the brane second fundamental form
as follows
\begin{subequations}\label{sff-}
\begin{eqnarray}
- K_{i \mu \nu} v^\mu dz^\nu &\equiv& n^a_i(z^\gamma + dz^\gamma)
V^\Sigma_{|| a}(z^\gamma + dz^\gamma) \notag \\ &\simeq& e_{c
\mu}(z^\gamma) v^\mu(z^\gamma) \partial_\nu n^c_i(z^\gamma) dz^\nu
+ n^a_i(z^\gamma) \Gamma^c_{\sps ab} e_{c \mu}(z^\gamma)
v^\mu(z^\gamma) e^b_\nu(z^\gamma) dz^\nu \notag \\ &=& g_{cd}
e^d_\mu \left(\partial_\nu n^c_i +  \Gamma^c_{\sps ab} n^a_i
e^b_\nu \right) v^\mu dz^\nu = g_{cd} e^d_\mu \ecnabla{\nu} n^c_i
\; v^\mu dz^\nu \label{sff} \\ &=& -n^c_i \left(\partial_\nu e_{c
\mu} - \Gamma^a_{\sps cb} e_{a \mu} e^b_\nu \right) v^\mu dz^\nu =
- g_{cd} n^d_i \ecnabla{\nu} e^c_\mu \; v^\mu dz^\nu. \label{sff1}
\end{eqnarray}
\end{subequations}
As above, let us perform the parallel transport of normal vector
to the brane world volume $V_a^\bot$ taking into account the bulk
connection $\Gamma^a_{\sps bc}$
\begin{equation}\label{s3}
V^\bot_{|| a}(z^\gamma + dz^\gamma) = V^\bot_a(z^\gamma) +
\Gamma^c_{\sps ab} V^\bot_c(z^\gamma) dx^b = V^\bot_a(z^\gamma) +
e^b_\lambda(z^\gamma) \Gamma^c_{\sps ab} V^\bot_c(z^\gamma)
dz^\lambda.
\end{equation}
The reader will have no difficulty in showing that the projections
of (\ref{s3}) in the basis $(n^a_i; e^a_\mu)$ defines the brane
connection $\dimi{\Sigma}\Gamma^j_{\sps i\lambda} \equiv g_{cd}
n^d_k q^{kj} \ecnabla{\lambda} n^c_i$ and the second fundamental
form $K_{i \mu \nu}$ exactly as (\ref{sff-}). Notice that
$\dimi{\Sigma} \Gamma^\lambda_{\sps \mu \nu}$,
$\dimi{\Sigma}\Gamma^j_{\sps i\lambda}$, and $K^i_{\sps \mu \nu}$
being the projections of $\ecnabla{\nu} e^a_\mu$, $\ecnabla{\nu}
n^a_i$ can be related via Gauss-Weingarten like equations as
follows
\begin{subequations}\label{g-w}
\begin{eqnarray}
\ecnabla{\nu} e^a_\mu &=& g_{cd} g^{da} \ecnabla{\nu} e^c_\mu =
g_{cd} \left(e^{d \lambda} e^a_\lambda + n^{di} n^a_i\right)
\ecnabla{\nu} e^c_\mu = \dimi{\Sigma} \Gamma^\lambda_{\sps \mu
\nu} e^a_\lambda + K^i_{\sps \mu \nu} n^a_i; \label{g-w1} \\
\ecnabla{\nu} n^a_i &=& g_{cd} g^{da} \ecnabla{\nu} n^c_i = g_{cd}
\left(e^{d \lambda} e^a_\lambda + n^{dj} n^a_j\right)
\ecnabla{\nu} n^c_i = \dimi{\Sigma} \Gamma^j_{\sps i \nu} n^a_j -
K^{\sps \lambda}_{i \dsps \nu} e^a_\lambda. \label{g-w2}
\end{eqnarray}
\end{subequations}
Motivated by the ideology of this paper we consider only the
totally antisymmetric part of the contortion understood as result
of a covariant splitting into irreducible parts (\ref{irrparts});
then $\Gamma_{abc} = \esi{\Gamma_{abc}} + \Theta_{abc}$ such that
$\Theta_{abc} = \Theta_{[abc]}$. Our next step is to decompose the
brane connection and the second fundamental form into Riemannian
and non-Riemannian parts:
\begin{equation}\label{s4}
\ecnabla{\nu} e^a_\mu = \enabla{\nu} e^a_\mu + \Theta^a_{\sps dc}
e^d_\mu e^c_\nu \equiv \enabla{\nu} e^a_\mu + \Theta^a_{\sps
\mu\nu};
\end{equation}
then the brane and the orthogonal projection of (\ref{g-w1})
becomes
\begin{subequations}\label{decomp}
\begin{eqnarray}
e^\gamma_a \ecnabla{\nu} e^a_\mu &=& \dimi{\Sigma}
\Gamma^\gamma_{\sps \mu \nu} = e^\gamma_a \enabla{\nu} e^a_\mu +
e^\gamma_a \Theta^a_{\sps \mu\nu} \equiv \dimi{\Sigma}
\esi{\Gamma^\gamma_{\sps \mu \nu}} + \; \Theta^\gamma_{\sps \mu
\nu}, \label{decomp1} \\ n^j_a \ecnabla{\nu} e^a_\mu &=& K^j_{\sps
\mu \nu} = n^j_a \enabla{\nu} e^a_\mu + n^j_a \Theta^a_{\sps
\mu\nu} \equiv \; \esi{K^j_{\sps \mu \nu}} + \; \Theta^j_{\sps \mu
\nu}. \label{decomp2}
\end{eqnarray}
\end{subequations}
Further, taking account of (\ref{decomp}) and (\ref{s4}) one can
express (\ref{g-w1}) in the form
\begin{equation}\label{s5}
\enabla{\nu} e^a_\mu + \Theta^a_{\sps \mu\nu} = \left(
\dimi{\Sigma} \esi{\Gamma^\gamma_{\sps \mu \nu}} + \;
\Theta^\gamma_{\sps \mu \nu} \right) e^a_\gamma + \left(
\esi{K^j_{\sps \mu \nu}} + \; \Theta^j_{\sps \mu \nu} \right)
n^a_j.
\end{equation}
The reader will easily prove that $e^a_\lambda
\dimi{\Sigma}\esi{\Gamma^\lambda_{\sps \mu\nu}}= e^a_\lambda
\dimi{\Sigma}\esi{\Gamma^\lambda_{\sps \nu\mu}}$ and similarly
$n^a_j \esi{K^j_{\sps \mu\nu}} = n^a_j \esi{ K^j_{\sps \nu\mu}}$;
then acting via symmetrization/antisymmetrization on (\ref{s5}) it
can easily be checked that
\begin{eqnarray}
\enabla{\nu} e^a_\mu = \dimi{\Sigma} \esi{\Gamma^\gamma_{\sps \mu
\nu}} e^a_\gamma + \esi{K^j_{\sps \mu \nu}} n^a_j; \quad
\Theta^a_{\sps \mu\nu} = \Theta^\gamma_{\sps \mu \nu} e^a_\gamma +
\; \Theta^j_{\sps \mu \nu} n^a_j.
\end{eqnarray}
Considering the definition of bulk covariant derivative along the
brane world volume direction one is able to define the brane
covariant derivative as follows
\begin{subequations}\label{s6}
\begin{eqnarray}
\dimi{\Sigma} \ecnabla{\nu} v_\mu &\equiv& \partial_\nu v_\mu -
\dimi{\Sigma} \Gamma^\lambda_{\sps \mu\nu} v_\lambda =
\partial_\nu (e^a_\mu V_a) - e_c^\lambda e^a_\lambda V_a
\ecnabla{\nu} e^c_\mu  = e^a_\mu \ecnabla{\nu} V_a + v_i K^i_{\sps
\mu\nu}, \label{s6a} \\ \dimi{\Sigma} \ecnabla{\nu} v_i &\equiv&
\partial_\nu v_i - \dimi{\Sigma} \Gamma^j_{\sps i \nu} v_j =
\partial_\nu (n^a_i V_a) - n^j_c n^a_j V_a \ecnabla{\nu} n^c_i =
n^a_i \ecnabla{\nu} V_a - v_\lambda K_{i \dsps \nu}^{\sps
\lambda}. \label{s6b}
\end{eqnarray}
\end{subequations}
Further, considering the brane covariant derivative we can relate
the bulk curvature tensor with the brane one. First notice a
definition
\begin{subequations}\label{s10}
\begin{eqnarray}
\ecnabla{[a} \ecnabla{b]} V_c &=& \partial_{[a} \ecnabla{b]} V_c -
\Gamma^d_{\sps [ba]} \ecnabla{d} V_c - \Gamma^d_{\sps c[a}
\ecnabla{b]} V_d \notag = \left( \partial_{[b}\Gamma^d_{\sps |c|
a]} + \Gamma^d_{\sps e[b} \Gamma^e_{\sps |c| a]} \right) V_d -
\Theta^d_{\sps ba} \ecnabla{d} V_c \notag \\ &\equiv& \frac{1}{2}
R^d_{\sps cba} V_d - \Theta^d_{\sps ba} \ecnabla{d} V_c, \quad
\text{and} \label{s10a} \\ \ecnabla{[a} \ecnabla{b]} V^c &=&
\frac{1}{2} R^{\sps c}_{d \sps ba} V^d - \Theta^d_{\sps ba}
\ecnabla{d} V^c. \label{s10b}
\end{eqnarray}
\end{subequations}
This definition gives a possibility to derive the following
properties of the curvature\footnote{These properties can be
checked by the decomposition: $R_{abcd} = \es{R}_{abcd} +
\widetilde{R}_{abcd}$.}
\begin{eqnarray}
&& \hspace{2cm} R_{abcd} = -R_{bacd} = -R_{abdc}, \notag \\ &&
R_{fabc} + R_{fbca} + R_{fcab} = -6 \left( \ecnabla{[a}
\Theta_{bc]f} + 2 \Theta^{d}_{\sps [ab} \Theta_{c]df} \right).
\notag
\end{eqnarray}
Now evaluate the commutator $e_a^\gamma \ecnabla{[\mu}
\ecnabla{\nu]} e_\alpha^a$
\begin{eqnarray}\label{s11}
e_a^\gamma \ecnabla{[\mu} \ecnabla{\nu]} e_\alpha^a &=& e_a^\gamma
e_{[\mu}^c \ecnabla{c}\left( \dimi{\Sigma} \Gamma^\lambda_{\sps
|\alpha| \nu]} e_\lambda^a + K^i_{\sps |\alpha| \nu]} n^a_i
\right) = \partial_{[\mu} \dimi{\Sigma} \Gamma^\gamma_{\sps
|\alpha| \nu]} + \dimi{\Sigma} \Gamma^\gamma_{\sps \lambda [\mu}
\dimi{\Sigma} \Gamma^\lambda_{\sps |\alpha| \nu]} - K^{i
\gamma}_{\dsps [\mu} K_{i |\alpha| \nu]} \notag
\\ &\equiv& \frac{1}{2} \left( \dimi{\Sigma} R^\gamma_{\sps \alpha
\mu \nu} - K_{i \alpha\nu} K^{i \gamma}_{\dsps \mu} + K_{i \alpha
\mu} K^{i \gamma}_{\dsps \nu} \right); \notag \\ e_a^\gamma
\ecnabla{[\mu} \ecnabla{\nu]} e_\alpha^a &=& e_a^\gamma \left(
e^c_\mu e^d_\nu \ecnabla{[c} \ecnabla{d]} e^a_\alpha +
\Theta^d_{\sps \nu \mu} \ecnabla{d} e^a_\alpha \right) =
\fracmy{1}{2} e_a^\gamma R^a_{\sps bcd}\: e^b_\alpha e^c_\mu
e^d_\nu. \notag
\end{eqnarray}
Therefore the Gauss like equation can be expressed as follows
\begin{equation}\label{g-eqs}
R^a_{\sps bcd} e^\gamma_a e^b_\alpha e^c_\mu e^d_\nu =
\dimi{\Sigma} R^\gamma_{\sps \alpha \mu \nu} - K_{i \alpha\nu}
K^{i \gamma}_{\dsps \mu} + K_{i \alpha \mu} K^{i \gamma}_{\dsps
\nu}.
\end{equation}
Finally, evaluate the commutator $n_a^j \ecnabla{[\mu}
\ecnabla{\nu]} e_\alpha^a$
\begin{eqnarray}
n^j_a \ecnabla{[\mu} \ecnabla{\nu]} e^a_\alpha &=& n^j_a
e^c_{[\mu} \ecnabla{c} \left( \dimi{\Sigma} \Gamma
^{\lambda}_{\sps |\alpha| \nu]} e^a_\lambda + K^{i}_{\sps |\alpha|
\nu]} n^a_i \right) = \fracmy{1}{2} \left( \dimi{\Sigma}
\ecnabla{\mu} K^j_{\sps \alpha \nu} - \dimi{\Sigma} \ecnabla{\nu}
K^j_{\sps \alpha \mu} - 2\Theta^{\lambda}_{\sps \mu \nu} K^j_{\sps
\alpha \lambda} \right); \notag \\ n^j_a \ecnabla{[\mu}
\ecnabla{\nu]} e^a_\alpha &=& n^j_a \left( e^c_\mu e^d_\nu
\ecnabla{[c} \ecnabla{d]} e^a_\alpha + \Theta^d_{\sps \nu \mu}
\ecnabla{d} e^a_\alpha \right) = \fracmy{1}{2} n^j_a R^{a}_{\sps
bcd} e^b_\alpha e^c_\mu e^d_\nu. \notag
\end{eqnarray}
Therefore the Codacci like equation can be expressed as follows
\begin{equation}\label{c-eqs}
R^{a}_{\sps bcd} \: n^i_a e^b_\alpha e^c_\mu e^d_\nu =
\dimi{\Sigma} \ecnabla{\mu} K^i_{\sps \alpha \nu} - \dimi{\Sigma}
\ecnabla{\nu} K^i_{\sps \alpha \mu} - 2\Theta^{\lambda}_{\sps \mu
\nu} K^i_{\sps \alpha \lambda}.
\end{equation}

\section{Field Equations in The Bulk}\label{app2}

The variational formalism of relativistic conservative medium with
internal degrees of freedom in the framework of EC space has been
investigated in the literature \cite{tors4, tors6, tors8}. The
dynamics of compactified type $IIB$ theory gives equations of
motion listed in the Appendix \ref{app3} [see
(\ref{eqs-motion-low})]. We modify these equations in order to
include the brane world contribution. Consider these modifications
at the level of action as follows
\begin{subequations}
\begin{eqnarray}\label{action-decomp}
&& \hspace{5cm} S_{total} = S + S_{c} + S_{br} + S_{cbr};
\\ S &=& \frac{1}{2k^2}\int d^{5}x \sqrt{|g|} \left[R - \fracmy{1}{2}
\left( \partial_a\phi \partial^a\phi + \fracmy{5}{8}
\partial_a\rho \partial^a\rho + \fracmy{5}{4} \partial_a\phi
\partial^a\rho + e^{2\phi + \fracmy{5}{4}\rho}\partial_a\chi
\partial^a\chi \right) \right. \notag \\ && \hspace{3cm} - \left.
\fracmy{1}{12} e^{-\phi} H_{abc}H^{abc} -
\fracmy{e^{\fracmy{5}{4}\rho}}{480} \widetilde{Y}_{abcde}
\widetilde{Y}^{abcde} + e^{-\fracmy{1}{2}\rho} \alpha \right],
\notag \\ S_{c} &=& \int{d^{5}x \sqrt{|g|} \left(- \Lambda
\right)}, \quad S_{cbr} = \int{d^{4}x \sqrt{|q|} \left( - \lambda
\right)}, \quad S_{br} = \int{d^{4}x \sqrt{|q|} \: L(q_{\mu \nu},
\psi_A, \dimi{4}\ecnabla{\mu}\psi_A)}, \notag \\ && \quad
\text{where} \hspace{1cm} \dimi{4} \ecnabla{\mu}\psi_A =
\partial_\mu \psi_A - \dimi{4} \Gamma^\alpha_{\sps \beta \mu}
\psi_A|_{\alpha}^{\sps \beta}; \quad \psi_A|_{\alpha}^{\sps \beta}
\equiv \phi_B \: \Omega^{B}_{\sps A}|_{\alpha}^{\sps \beta}.
\end{eqnarray}
\end{subequations}
The term $\Omega^{B}_{\sps A}|_{\alpha}^{\sps \beta}$ is the
generator of coordinate transformations and its explicit form
depends on the order of the field $\psi_A$. By construction, we
suppose $g_{ab} n^a n^b = 1$, $x^4 \equiv y$, and $e^a_\mu =
\delta^a_\mu$; then the formalism constructed in the Appendix
\ref{app1} can straightforwardly be applied. Further, we introduce
a notation by taking the following variation
\begin{eqnarray}
\delta \int{d^{5}x \sqrt{|g|} \dimi{5}R} &=& \delta \int{d^{5}x
\sqrt{|g|} \left(\dimi{5} \hspace{-0.15cm} \es{R} - \Theta^2
\right)} \notag \\ &=& \int{d^{5}x \sqrt{|g|} \left( \es{G}_{ab} +
\fracmy{1}{2} g_{ab} \Theta^2 - 3\Theta_{acd} \Theta_b^{\sps cd}
\right)} \delta g^{ab} \\ &=& \int{d^{5}x \sqrt{|g|} \: G_{(ab)}}
\delta g^{ab}, \quad \text{where} \notag \\ G_{ab} \equiv R_{ab} -
\fracmy{1}{2} g_{ab} \dimi{5} R &-& 2\Theta_{acd} \Theta_b^{\sps
cd}; \quad \Theta_{abc} = \fracmy{\sqrt{3}}{2}
e^{\fracmy{1}{2}\left(\phi + \fracmy{5}{4}\rho\right)} \left(
\partial_{[a}A_{bc]} - \chi \: \partial_{[a}B_{bc]} \right). \notag
\end{eqnarray}
The non-zero variations of $(S_{c} + S_{cbr})$ gives the following
expressions
\begin{subequations}
\begin{eqnarray}
\delta S_{c} &=& -\delta \int{d^{5}x \sqrt{|g|} \; \Lambda} =
\fracmy{1}{2} \int{d^{5}x \sqrt{|g|} \; g_{ab} \Lambda} \; \delta
g^{ab}, \\ \delta S_{cbr} &=& -\delta \int{d^{5}x \sqrt{|q|} \;
\lambda \: \delta(y)} = \fracmy{1}{2} \int{d^{5}x \sqrt{|q|} \:
\lambda \: q_{\mu \nu} \: \delta^\mu_a \delta^\nu_b \: \delta
g^{ab} \: \delta(y)}.
\end{eqnarray}
\end{subequations}
Variations of $S_{br}$ with respect to \{$g^{ab}, \rho, \phi,
\chi, A_{ab}, B_{ab}, \psi_A$\} gives respectively
\begin{subequations}
\begin{eqnarray}
\delta S_{br} &=& \int d^4 x \sqrt{|q|} \: \left[
\fracmy{1}{\sqrt{|q|}} \fracmy{\partial(\sqrt{|q|} L)}{\partial
q_{\alpha \beta}} \delta q_{\alpha \beta} - \fracmy{\partial
L}{\partial \dimi{4} \ecnabla{\lambda} \psi_A} \psi_A |^{\sps
\beta}_{\alpha} \delta \Gamma^\alpha_{\sps \beta \mu} \right]
\\ &\equiv& -\fracmy{1}{2} \int d^5 x \sqrt{|q|} \:
\tau_{\alpha \beta} \: \delta_a^\alpha \delta_b^\beta \delta(y)
\delta g^{ab} = -\fracmy{1}{2} \int d^5 x \sqrt{|q|} \: \left(
\es{\tau}_{\alpha \beta} + \: \widetilde{\tau}_{\alpha \beta}
\right) \: \delta_a^\alpha \delta_b^\beta \delta(y) \delta g^{ab},
\notag
\end{eqnarray}
\begin{eqnarray}
\delta S_{br} = -\fracmy{5}{32\sqrt{3}} \int d^{5}x \sqrt{|q|} \:
e^{\fracmy{1}{2}\left(\phi + \fracmy{5}{4}\rho \right)}
\theta^{\mu \alpha \nu} \: \widetilde{F}_{\alpha \mu \nu} \:
\delta(y) \: \delta \rho,
\end{eqnarray}
\begin{eqnarray}
\delta S_{br} = -\fracmy{1}{8\sqrt{3}} \int d^{5}x \sqrt{|q|} \:
e^{\fracmy{1}{2}\left(\phi + \fracmy{5}{4}\rho \right)}
\theta^{\mu \alpha \nu} \: \widetilde{F}_{\alpha \mu \nu} \:
\delta(y) \: \delta \phi,
\end{eqnarray}
\begin{eqnarray}
\delta S_{br} = -\fracmy{1}{4\sqrt{3}} \int d^{5}x \sqrt{|q|} \:
e^{\fracmy{1}{2}\left(\phi + \fracmy{5}{4}\rho \right)}
\theta^{\alpha \mu \nu} \: H_{\alpha \mu \nu} \: \delta(y) \:
\delta \chi,
\end{eqnarray}
\begin{eqnarray}
\delta S_{br} &=& \fracmy{1}{4\sqrt{3}} \int{d^{4}x \sqrt{|q|} \:
e^{\fracmy{1}{2}\left(\phi + \fracmy{5}{4}\rho \right)}
\theta^{\alpha \mu \nu} \: \delta F_{\alpha \mu \nu}} \\ &=&
-\fracmy{\sqrt{3}}{4} \int d^{5}x \sqrt{|q|} \;
e^{\fracmy{1}{2}\left(\phi + \fracmy{5}{4}\rho \right)} \left[
\sigma^{\mu \nu} + \fracmy{1}{2} \theta^{\alpha \mu \nu}
\partial_\alpha \left( \phi + \fracmy{5}{4}\rho \right) \right]
\delta^a_\mu \delta^b_\nu \delta(y) \delta A_{ab}, \notag
\end{eqnarray}
\begin{eqnarray}
\delta S_{br} &=& -\fracmy{1}{4\sqrt{3}} \int{d^{4}x \sqrt{|q|} \:
e^{\fracmy{1}{2}\left(\phi + \fracmy{5}{4}\rho \right)} \chi \:
\theta^{\alpha \mu \nu} \: \delta H_{\alpha \mu \nu}} \\ &=&
-\fracmy{\sqrt{3}}{4} \int d^{5}x \sqrt{|q|} \;
e^{\fracmy{1}{2}\left(\phi + \fracmy{5}{4}\rho \right)} \left[
\chi \: \sigma^{\nu \mu} + \theta^{\alpha \nu \mu} \left(
\partial_\alpha \chi + \fracmy{1}{2} \chi \partial_\alpha \phi
+ \fracmy{5}{8} \chi \partial_\alpha \rho  \right) \right]
\delta^a_\mu \delta^b_\nu \delta(y) \delta B_{ab}, \notag
\end{eqnarray}
\begin{eqnarray}\label{some11}
\fracmy{\delta S_{br}}{\delta \psi_A} = \int d^4 x \sqrt{|q|}
\left( \fracmy{\partial L}{\partial \psi_A} - \dimi{4}
\ecnabla{\lambda} \fracmy{\partial L}{\partial \dimi{4}
\ecnabla{\lambda} \psi_A} \right) = 0,
\end{eqnarray}
\end{subequations}
\begin{equation}
\text{where} \hspace{1cm} \sigma_{\mu \nu} \equiv \dimi{4}
\enabla{\alpha} \theta^\alpha_{\sps \mu \nu}, \quad \theta_{\alpha
\beta \lambda} \equiv \xi_{[\alpha \beta \lambda]}, \quad
\xi_{\alpha}^{\sps \beta \lambda} \equiv 2\fracmy{\partial
L}{\partial \dimi{4}\Gamma^{\alpha}_{\sps \beta \lambda}} = -
2\fracmy{\partial L}{\partial \dimi{4} \ecnabla{\lambda} \psi_A}
\phi_B \: \Omega^B_{\sps A}|_{\alpha}^{\sps \beta}, \notag
\end{equation}
and $\{ \es{\tau}_{\mu \nu}, \widetilde{\tau}_{\mu \nu} \}$
tensors have generated by $\{ \es{L}, \widetilde{L} \}$ parts of
the brane Lagrangian [see (\ref{emt-decomp})]. \newline Due to
modifications given by (\ref{action-decomp}) the equations of
motion (\ref{eqs-motion-low}) become
\begin{subequations}\label{f-eqs}
\begin{eqnarray}\label{f-eqs1}
G_{(ab)} &=& k^2 T_{ab}; \quad T_{ab} = - g_{ab} \Lambda + S_{\mu
\nu} \delta_a^\mu \delta_b^\nu \: \delta(y) + k^{-2} P_{ab}, \quad
\text{where} \\ P_{ab} &=& \fracmy{1}{2} \partial_a \phi
\partial_b \phi + \fracmy{5}{8} \partial_{(a} \phi \partial_{b)} \rho +
\fracmy{5}{16} \partial_a \rho \partial_b \rho  + \fracmy{e^{2\phi
+ \fracmy{5}{4}\rho}}{2} \partial_a \chi \partial_b \chi +
\fracmy{e^{-\phi}}{4} H_{acd} H_{b}^{\sps cd} \notag \\ &+&
\fracmy{e^{\fracmy{5}{4}\rho}}{96} \widetilde{Y}_{acdef}
\widetilde{Y}_{b}^{\sps cdef} - \fracmy{1}{2}g_{ab} \left(
\fracmy{1}{2} \partial_c \phi \partial^c \phi + \fracmy{5}{8}
\partial_c \phi \partial^c \rho + \fracmy{5}{16} \partial_c \rho
\partial^c \rho \right. \notag \\ &+& \left. \fracmy{e^{2\phi
+ \fracmy{5}{4} \rho}}{2} \partial_c \chi \partial^c \chi +
\fracmy{e^{-\phi}}{12} H^2 - e^{-\fracmy{\rho}{2}} \alpha \right);
\quad S_{\mu \nu} \equiv \left( -q_{\mu \nu} \lambda + \tau_{\mu
\nu} \right), \notag
\end{eqnarray}
\begin{eqnarray}\label{f-eqs2}
&& \dimi{5}\enabla{c} \hspace{-0.2cm} \dimi{5}\uenabla{c} \rho +
\dimi{5}\enabla{c} \hspace{-0.2cm} \dimi{5}\uenabla{c} \phi -
e^{\fracmy{5}{4}\rho} \left( e^{2\phi} \partial_a \chi
\partial^a \chi + \fracmy{e^\phi}{6} \widetilde{F}^2 + \fracmy{1}{120}
\widetilde{Y}^2 \right) - \fracmy{4}{5} e^{-\fracmy{1}{2}\rho}
\alpha \notag \\ &=& \fracmy{k^2}{2\sqrt{3}}
e^{\fracmy{1}{2}\left(\phi + \fracmy{5}{4}\rho \right)}
\theta^{\mu \alpha \nu} \: \widetilde{F}_{\alpha \mu \nu} \:
\delta(y),
\end{eqnarray}
\begin{eqnarray}\label{f-eqs3}
&& \fracmy{e^{-\phi}}{12} H^2 - e^{2\phi + \fracmy{5}{4}\rho}
\partial_a \chi \partial^a \chi - \fracmy{e^{\phi
+ \fracmy{5}{4}\rho}}{12} \widetilde{F}^2 + \dimi{5}\enabla{a}
\hspace{-0.2cm} \dimi{5}\uenabla{a} \phi + \fracmy{5}{8}
\dimi{5}\enabla{a} \hspace{-0.2cm} \dimi{5}\uenabla{a} \rho \notag
\\ &=& \fracmy{k^2}{4\sqrt{3}} e^{\fracmy{1}{2}\left(\phi +
\fracmy{5}{4}\rho \right)} \theta^{\mu \alpha \nu} \:
\widetilde{F}_{\alpha \mu \nu} \: \delta(y),
\end{eqnarray}
\begin{eqnarray}\label{f-eqs4}
\dimi{5} \uenabla{a} \left( e^{2\phi + \fracmy{5}{4}\rho}
\partial_a \chi \right) + \fracmy{e^{\phi + \fracmy{5}{4}\rho}}{6}
\widetilde{F}_{abc}H^{abc} = \fracmy{k^2}{2\sqrt{3}} \:
e^{\fracmy{1}{2}\left(\phi + \fracmy{5}{4}\rho \right)}
\theta^{\alpha \mu \nu} \: H_{\alpha \mu \nu} \: \delta(y),
\end{eqnarray}
\begin{eqnarray}\label{f-eqs5}
&& 24 \dimi{5} \enabla{c} \left( e^{\phi + \fracmy{5}{4}\rho}
\widetilde{F}^{cab} \right) + e^{\fracmy{5}{4}\rho}
\widetilde{Y}^{abcde} H_{cde} + 3 \dimi{5} \enabla{c} \left(
e^{\fracmy{5}{4}\rho} \widetilde{Y}^{cdeab} B_{de} \right) \notag
\\ &=& 12 \sqrt{3} \: k^2 \: e^{\fracmy{1}{2}\left(\phi +
\fracmy{5}{4}\rho \right)} \left[ \theta^{\alpha \mu \nu}
\partial_\alpha \left( \phi + \fracmy{5}{4} \rho \right) + 2
\sigma^{\mu \nu} \right] \delta^a_\mu \delta^b_\nu \: \delta(y),
\end{eqnarray}
\begin{eqnarray}\label{f-eqs6}
&& 24 \dimi{5} \enabla{c} \left( e^{-\phi} H^{cab} -  e^{\phi +
\fracmy{5}{4}\rho} \chi \widetilde{F}^{cab} \right) - 3\dimi{5}
\enabla{c} \left( e^{\fracmy{5}{4}\rho} \widetilde{Y}^{decab}
A_{de} \right) - e^{\fracmy{5}{4}\rho} \widetilde{Y}^{abcde}
F_{cde} \notag \\ && = 24 \sqrt{3} \: k^2 \:
e^{\fracmy{1}{2}\left(\phi + \fracmy{5}{4}\rho \right)} \left[
\theta^{\alpha \nu \mu} \left( \partial_\alpha \chi +
\fracmy{1}{2} \chi \: \partial_\alpha \phi + \fracmy{5}{8} \chi \:
\partial_\alpha \rho  \right) + \chi \: \sigma^{\nu \mu} \right]
\delta^a_\mu \delta^b_\nu \: \delta(y),
\end{eqnarray}
\begin{eqnarray}\label{f-eqs7}
\widetilde{Y}^{abcde} = -\varrho \: \eta^{abcde} \:
e^{-\fracmy{5}{4}\rho}; \quad \widetilde{Y}_{abcde} = - \varrho \:
\eta_{abcde} \: e^{\fracmy{5}{4}\rho},
\end{eqnarray}
\begin{eqnarray}\label{f-eqs8}
\fracmy{\partial L}{\partial \psi_A} - \dimi{4} \ecnabla{\lambda}
\fracmy{\partial L}{\partial \dimi{4} \ecnabla{\lambda} \psi_A} =
0.
\end{eqnarray}
\end{subequations}
The relations (\ref{f-eqs7}) are the consequences of self-duality
conditions, the ansatz (\ref{ansatz}), and the equation of motion
(\ref{eqs-motion-low8}). Hence throughout the paper we consider
(\ref{f-eqs7}) instead of (\ref{eqs-motion-low8}). On the other
hand, on the brane survive only $\widetilde{Y}_{\alpha \beta
\gamma \lambda 4}$ components. It can be checked that
\begin{eqnarray}\label{some9}
\widetilde{Y}^{\alpha \beta \gamma \lambda 4} &=& -\sqrt{|q|} \:
\eta^{\alpha \beta \gamma \lambda} \: \widetilde{Y}^{01234} =
-\varrho \: \eta^{\alpha \beta \gamma \lambda} \:
e^{-\fracmy{5}{4}\rho}, \notag \\ \widetilde{Y}_{\alpha \beta
\gamma \lambda 4} &=& \fracmy{1}{\sqrt{|q|}} \: \eta_{\alpha \beta
\gamma \lambda} \: \widetilde{Y}_{01234} = - \varrho \:
\eta_{\alpha \beta \gamma \lambda} \: e^{\fracmy{5}{4}\rho},
\notag \\ \text{and} \quad \widetilde{Y}_{\alpha \beta \gamma
\lambda 4} \widetilde{Y}^{\alpha \beta \gamma \lambda 4} &=& - 4!
\: \varrho^2, \quad \text{where} \quad \eta_{\alpha \beta \gamma
\lambda} = \eta_{abcde} e_\alpha^a e_\beta^b e_\gamma^c
e_\lambda^d n^e
\end{eqnarray}
is the brane volume form and $\varrho$ is the Freund-Rubin
parameter.

\section{Equations of Motion for The Type {\bf IIB} String}\label{app3}

We have noticed that the massless bosonic sector of the type $IIB$
string can not be described by a covariant action \cite{str17}
however the covariant equations of motion exists. The
10-dimensional equations considered in this paper are as follows
\begin{subequations}\label{eqs-motion}
\begin{eqnarray}\label{eqs-motion1}
0 &=& \es{\widetilde{R}}_{AB} + \fracmy{5}{4} \enabla{(A}
\enabla{B)} \widetilde{\Phi} - \fracmy{e^{-\Phi}}{4}
H_{ACD}H_{B}^{\sps CD} - \fracmy{1}{2}
\partial_A \Phi \partial_B \Phi - \fracmy{5}{8} \partial_{(A} \Phi
\partial_{B)} \widetilde{\Phi} - \fracmy{e^{2\Phi +
\fracmy{5}{4}\widetilde{\Phi}}}{2} \partial_A \Psi \partial_B \Psi
\notag \\ &-& \fracmy{e^{\Phi + \fracmy{5}{4}\widetilde{\Phi}}}{4}
\widetilde{F}_{ACD} \widetilde{F}_{B}^{\sps CD} -
\fracmy{e^{\fracmy{5}{4}\widetilde{\Phi}}}{96}
\widetilde{Y}_{ACDEF} \widetilde{Y}_{B}^{\sps CDEF} -
\fracmy{1}{2}\widetilde{G}_{AB} \left(\es{\widetilde{R}} +
\fracmy{5}{2} \enabla{C} \uenabla{C} \widetilde{\Phi} -
\fracmy{e^{-\Phi}}{12} H^2 \right. \notag \\ &-& \left.
\fracmy{1}{2} \partial_C \Phi \partial^C \Phi - \fracmy{5}{8}
\partial_C \Phi \partial^C \widetilde{\Phi} - \fracmy{25}{16} \partial_C
\widetilde{\Phi} \partial^C \widetilde{\Phi} - \fracmy{e^{2\Phi +
\fracmy{5}{4}\widetilde{\Phi}}}{2} \partial_C \Psi
\partial^C \Psi -  \fracmy{e^{\Phi +
\fracmy{5}{4}\widetilde{\Phi}}}{12} \widetilde{F}^2 \right),
\end{eqnarray}
\begin{eqnarray}\label{eqs-motion2}
0 = \es{\widetilde{R}} - \fracmy{e^{-\Phi}}{12} H^2 -
\fracmy{1}{2} \partial_A \Phi \partial^A \Phi - \fracmy{25}{16}
\partial_A \widetilde{\Phi} \partial^A \widetilde{\Phi} - \fracmy{1}{2}
\enabla{A} \uenabla{A} \Phi + \fracmy{5}{2} \enabla{A} \uenabla{A}
\widetilde{\Phi},
\end{eqnarray}
\begin{eqnarray}
0 &=& \fracmy{e^{-\Phi}}{12} H^2 - \fracmy{5}{4} \partial_A
\widetilde{\Phi} \partial^A \Phi - \fracmy{25}{32} \partial_A
\widetilde{\Phi} \partial^A \widetilde{\Phi} - e^{2\Phi +
\fracmy{5}{4}\widetilde{\Phi}} \partial_A \Psi
\partial^A \Psi \notag \\ && \hspace{5cm} - \fracmy{e^{\Phi
+ \fracmy{5}{4}\widetilde{\Phi}}}{12} \widetilde{F}^2 + \enabla{A}
\uenabla{A} \Phi + \fracmy{5}{8} \enabla{A} \uenabla{A}
\widetilde{\Phi},
\end{eqnarray}
\begin{equation}
0 = \enabla{A} \uenabla{A} \Psi + 2 \partial_A \Psi \partial^A
\Phi + \fracmy{e^{-\Phi}}{6} \widetilde{F}_{ABC} H^{ABC},
\end{equation}
\begin{eqnarray}
0 &=& 24 \: e^\Phi \left( \partial_C \Phi \: \widetilde{F}^{CAB} +
\enabla{C} \widetilde{F}^{CAB} \right) + \widetilde{Y}^{ABCDE}
H_{CDE} + 3 \enabla{C} \left( \widetilde{Y}^{CDEAB} B_{DE}
\right), \quad
\end{eqnarray}
\begin{eqnarray}
0 &=& 24 \: e^{-\fracmy{5}{4} \widetilde{\Phi} - \Phi} \left[
\partial_C \left(\Phi + \fracmy{5}{4}\widetilde{\Phi} \right)
H^{CAB} - \enabla{C} H^{CAB} \right] + 3\enabla{C} \left(
\widetilde{Y}^{DECAB} A_{DE} \right)  \notag \\ &+& 24 \: e^{\Phi}
\left[ \enabla{C} \left( \Psi \widetilde{F}^{CAB} \right) + \Psi
\widetilde{F}^{CAB} \partial_C \Phi \right] +
\widetilde{Y}^{ABCDE} F_{CDE},
\end{eqnarray}
\begin{eqnarray}
0 = \enabla{E} \widetilde{Y}^{EABCD},
\end{eqnarray}
and the imposed self-duality condition
\begin{eqnarray}\label{self-duality}
\widetilde{Y}_{ABCDE} = *\widetilde{Y}_{ABCDE} &=& \fracmy{1}{5!}
\eta_{J_1 J_2 J_3 J_4 J_5 ABCDE} \widetilde{Y}^{J_1 J_2 J_3 J_4
J_5}.
\end{eqnarray}
\end{subequations}
We observe that these equations can be derived from the following
action
\begin{eqnarray}\label{string-eff3}
S_E = \frac{1}{2k^2_{10}}\int d^{10}X \sqrt{|\widetilde{G}|} \:
e^{-\fracmy{5}{4}\widetilde{\Phi}} \hspace{-0.4cm} && \left[
\es{\widetilde{R}} - \fracmy{e^{-\Phi}}{12}H^2 - \fracmy{1}{2}
\partial_A \Phi \partial^A \Phi - \fracmy{5}{8}\partial_A \Phi
\partial^A \widetilde{\Phi} + \fracmy{25}{16} \partial_A \widetilde{\Phi} \partial^A
\widetilde{\Phi}  \right. \notag \\ &&  \left. - \fracmy{e^{2\Phi
+ \fracmy{5}{4}\widetilde{\Phi}}}{2} \partial_A \Psi \partial^A
\Psi - \fracmy{e^{\Phi + \fracmy{5}{4}\widetilde{\Phi}}}{12}
\widetilde{F}^2 - \fracmy{e^{\fracmy{5}{4}\widetilde{\Phi}}}{480}
\widetilde{Y}^2 \right];
\end{eqnarray}
\begin{eqnarray}
H_{ABC} &=& 3\partial_{[A}B_{BC]}, \quad F_{ABC} =
3\partial_{[A}A_{BC]}, \quad \widetilde{F}_{ABC} = F_{ABC} - \Psi
H_{ABC}; \notag \\ Y_{ABCDE} &=& 5\partial_{[A}W_{BCDE]}, \quad
\widetilde{Y}_{ABCDE} = Y_{ABCDE} - 5 A_{[AB}H_{CDE]} + 5
B_{[AB}F_{CDE]}; \notag
\end{eqnarray}
except for vanishing of the term $\widetilde{Y}^2$ in
(\ref{eqs-motion1}) and the self-duality condition
(\ref{self-duality}).

Applying the method \cite{str12} based on the ansatz
(\ref{ansatz}) one yields 5-dimensional equations of motion in the
form
\begin{subequations}\label{eqs-motion-low}
\begin{eqnarray}\label{eqs-motion-low1}
0 &=& \dimi{5} \es{R}_{ab} - \fracmy{e^{-\phi}}{4}
H_{acd}H_{b}^{\sps cd} - \fracmy{1}{2}
\partial_a \phi \partial_b \phi - \fracmy{5}{8} \partial_{(a} \phi
\partial_{b)} \rho - \fracmy{5}{16} \partial_a \rho \partial_b \rho  -
\fracmy{e^{2\phi + \fracmy{5}{4}\rho}}{2} \partial_a \chi
\partial_b \chi \notag \\ &-& \fracmy{e^{\phi +
\fracmy{5}{4}\rho}}{4} \widetilde{F}_{acd} \widetilde{F}_{b}^{\sps
cd} - \fracmy{e^{\fracmy{5}{4}\rho}}{96} \widetilde{Y}_{acdef}
\widetilde{Y}_{b}^{\sps cdef} - \fracmy{1}{2}g_{ab} \left(
\dimi{5}\es{R} - \fracmy{e^{-\phi}}{12} H^2 - \fracmy{1}{2}
\partial_c \phi \partial^c \phi - \fracmy{5}{8} \partial_c \phi
\partial^c \rho \right. \notag \\ &-& \left. \fracmy{5}{16}
\partial_c \rho \partial^c \rho - \fracmy{e^{2\phi + \fracmy{5}{4} \rho}}{2}
\partial_c \chi \partial^c \chi -  \fracmy{e^{\phi + \fracmy{5}{4}\rho}}{12}
\widetilde{F}^2 + e^{-\fracmy{\rho}{2}} \: \dimi{\bot} R \right),
\end{eqnarray}
\begin{equation}\label{eqs-motion-low23}
0 = \dimi{5}\enabla{c} \hspace{-0.2cm} \dimi{5}\uenabla{c} \rho +
\dimi{5}\enabla{c} \hspace{-0.2cm} \dimi{5}\uenabla{c} \phi -
e^{\fracmy{5}{4}\rho} \left( e^{2\phi} \partial_a \chi
\partial^a \chi + \fracmy{e^\phi}{6} \widetilde{F}^2 +
\fracmy{1}{120} \widetilde{Y}^2 \right) - \fracmy{4}{5}
e^{-\fracmy{1}{2}\rho} \: \dimi{\bot} R.
\end{equation}
\begin{equation}\label{eqs-motion-low4}
0 = \fracmy{e^{-\phi}}{12} H^2 - e^{2\phi + \fracmy{5}{4}\rho}
\partial_a \chi \partial^a \chi - \fracmy{e^{\phi
+ \fracmy{5}{4}\rho}}{12} \widetilde{F}^2 + \dimi{5}\enabla{a}
\hspace{-0.2cm} \dimi{5}\uenabla{a} \phi + \fracmy{5}{8}
\dimi{5}\enabla{a} \hspace{-0.2cm} \dimi{5}\uenabla{a} \rho,
\end{equation}
\begin{equation}\label{eqs-motion-low5}
0 = \dimi{5} \enabla{a} \hspace{-0.2cm} \dimi{5}\uenabla{a} \chi +
2 \partial_a \chi \partial^a \phi + \fracmy{5}{4} \partial_a \chi
\partial^a \rho  + \fracmy{e^{-\phi}}{6} \widetilde{F}_{abc} H^{abc},
\end{equation}
\begin{eqnarray}\label{eqs-motion-low6}
0 &=& 24 \: e^\phi \left[ \partial_c \left( \phi + \fracmy{5}{4}
\rho \right) \widetilde{F}^{cab} + \dimi{5}\enabla{c}
\widetilde{F}^{cab} \right] + 3 \dimi{5} \enabla{c} \left(
\widetilde{Y}^{cdeab} B_{de} \right) \notag \\ &+&
\widetilde{Y}^{abcde}H_{cde} + \fracmy{15}{4} \partial_c \rho \:
\widetilde{Y}^{cdeab} B_{de} ,
\end{eqnarray}
\begin{eqnarray}\label{eqs-motion-low7}
0 &=& 24 \: e^{\phi} \left[ \dimi{5}\enabla{c} \left( \chi
\widetilde{F}^{cab} \right) + \chi \widetilde{F}^{cab} \partial_c
\left( \phi + \fracmy{5}{4} \rho \right) \right] + 24 \:
e^{-\fracmy{5}{4} \rho - \phi} \left[
\partial_c \phi H^{cab} - \dimi{5}\enabla{c} H^{cab} \right]
\notag \\ &+& 3\dimi{5}\enabla{c} \left( \widetilde{Y}^{decab}
A_{de} \right) + \fracmy{15}{4} \partial_c \rho \:
\widetilde{Y}^{decab} A_{de} + \widetilde{Y}^{abcde} F_{cde},
\end{eqnarray}
\begin{eqnarray}\label{eqs-motion-low8}
0 = \dimi{5}\enabla{e} \widetilde{Y}^{eabcd} + \fracmy{5}{4}
\partial_e \rho \: \widetilde{Y}^{eabcd},
\end{eqnarray}
\end{subequations}
where $\dimi{\bot} R = const.$ is the scalar curvature of $S^5$.
\newline We observe that these equations can be derived from the
following action
\begin{eqnarray}
S_0 = \frac{1}{2k^2}\int d^{5}x \sqrt{|g|} \hspace{-0.4cm} &&
\left[\es{R} - \fracmy{1}{2} \left( \partial_a\phi
\partial^a\phi + \fracmy{5}{8} \partial_a\rho \partial^a\rho +
\fracmy{5}{4} \partial_a\phi \partial^a\rho + e^{2\phi +
\fracmy{5}{4}\rho}\partial_a\chi \partial^a\chi \right) \right.
\notag \\ && - \left. \fracmy{1}{12} \left( e^{\phi +
\fracmy{5}{4}\rho} \widetilde{F}^2 + e^{-\phi}H^2 \right) -
\fracmy{e^{\fracmy{5}{4}\rho}}{480} \widetilde{Y}^2 +
e^{-\fracmy{1}{2}\rho} \: \dimi{\bot} R \right];
\end{eqnarray}
\begin{eqnarray}
F_{abc} &=& 3\partial_{[a}A_{bc]}, \quad H_{abc} =
3\partial_{[a}B_{bc]}, \quad \widetilde{F}_{abc} = F_{abc} - \chi
H_{abc}, \notag \\ Y_{abcde} &=& 5\partial_{[a}W_{bcde]}, \quad
\widetilde{Y}_{abcde} = Y_{abcde} - 5 A_{[ab}H_{cde]} + 5
B_{[ab}F_{cde]}, \notag
\end{eqnarray}
except for vanishing of the term $\widetilde{Y}^2$ in
(\ref{eqs-motion-low1}), for the wrong coefficient of
$\widetilde{Y}^2$ in (\ref{eqs-motion-low23}), and for
consequences of the self-duality condition given by (\ref{some8}).


\begin{thebibliography}{50}

 \bibitem{str1}
  C. Lovelace,
  Strings in curved space,
  Phys. Lett. B {\bf 135}, 75 (1984);
  E.S. Fradkin, A.A Tseytlin,
  Quantum string theory effective action,
  Nucl. Phys. B {\bf 261}, 1 (1985);
  C.G. Callan, D. Friedan, E.J. Martinec, M.J. Perry,
  Strings in background fields,
  Nucl. Phys. B {\bf 262}, 593 (1985).

 \bibitem{str2}
  T.L. Curtright, C.K. Zachos,
  Geometry, topology and supersymmetry in non-linear models,
  Phys. Rev. Lett. B {\bf 53}, 1799 (1984).

 \bibitem{str3}
  C.M. Hull,
  The Geometry of N=2 Strings with Torsion,
  Phys. Lett. B {\bf 387}, 497 (1996), \newline
  \href{http://www.arxiv.org/abs/hep-th/9606190}{hep-th/9606190}.

 \bibitem{str5}
  R.T. Hammond,
  Strings in gravity with torsion,
  Gen. Rel. Grav. {\bf 32}, 2007 (2000), \newline
  \href{http://www.arxiv.org/abs/gr-qc/9904033}{gr-qc/9904033}.

 \bibitem{str6}
  I.L. Shapiro,
  Physical Aspects of the Space-Time Torsion,
  Phys. Rept. {\bf 357}, 113 (2001), \newline
  \href{http://www.arxiv.org/abs/hep-th/0103093}{hep-th/0103093}.

 \bibitem{str11}
  J. Polchinski, \textit{String Theory},
  (Cambridge University Press, Cambridge 1998).

 \bibitem{str12}
  M. Sato, A. Tsuchiya,
  Hamilton-Jacobi Method and Effective Actions of D-brane and M-brane in
  Supergravity,
  Nucl. Phys. B {\bf 671}, 293 (2003),
  \href{http://www.arxiv.org/abs/hep-th/0305090}{hep-th/0305090}
  M. Sato, A. Tsuchiya,
  Born-Infeld Action from Supergravity,
  Prog. Theor. Phys. {\bf 109}, 687 (2003),
  \href{http://www.arxiv.org/abs/hep-th/0211074}{hep-th/0211074}.

 \bibitem{str13}
  H. Nastase, D. Vaman,
  On the nonlinear KK reductions on spheres of supergravity
  theories,
  Nucl. Phys. B {\bf 583}, 211 (2000),
  \href{http://www.arxiv.org/abs/hep-th/0002028}{hep-th/0002028};
  M. Cvetic, H. Lu, C. N. Pope,
  Consistent Kaluza-Klein Sphere Reductions,
  Phys. Rev. D {\bf 62}, 064028 (2000),
  \href{http://www.arxiv.org/abs/hep-th/0003286}{hep-th/0003286}.

 \bibitem{str14}
  A. A. Tseytlin,
  Ambiguity in the effective action in string theories,
  Phys. Lett. B {\bf 176}, 92 (1986);
  M. C. Bento, N. E. Mavromatos,
  Ambiguities in the low-energy effective actions of string
  theories with the inclusion of antisymmetric tensor and dilaton
  fields,
  Phys. Lett. B {\bf 190}, 105 (1987).

 \bibitem{str15}
  G. Horowitz and A. Strominger,
  Black Strings and p-branes,
  Nucl. Phys. B {\bf 360}, 197 (1991);
  T. Mohaupt,
  Black Holes in Supergravity and String Theory,
  Class. Quant. Grav. {\bf 17}, 3429 (2000),
  \href{http://www.arxiv.org/abs/hep-th/0004098}{hep-th/0004098}.

 \bibitem{str17}
  N. Marcus and J.H. Schwarz,
  Field theories that have no manifestly Lorentz-invariant
  formulation,
  Phys. Lett. B {\bf 115}, 111 (1982).

 \bibitem{str18}
  J.H. Schwarz,
  Covariant field equations of chiral N = 2 D = 10 supergravity,
  Nucl. Phys. B {\bf 226}, 269 (1983);
  P. Howe and P. West,
  The complete N = 2 D = 10 supergravity,
  Nucl. Phys. B {\bf 238}, 181 (1984).

 \bibitem{str19}
  M. Cvetic, M.J. Duff, James T. Liu, H. Lu, C.N. Pope, K.S.
  Stelle,
  Randall-Sundrum Brane Tensions,
  Nucl. Phys. B {\bf 605}, 141 (2001),
  \href{http://www.arxiv.org/abs/hep-th/0011167}{hep-th/0011167};
  M.J. Duff, James T. Liu, K.S. Stelle,
  A supersymmetric Type IIB Randall-Sundrum realization,
  J. Math. Phys. {\bf 42}, 3027 (2001),
  \href{http://www.arxiv.org/abs/hep-th/0007120}{hep-th/0007120};

 \bibitem{str20}
  M.S. Bremer, M.J. Duff, H. L\"{u}, C.N. Pope and K.S. Stelle,
  Instanton cosmology and domain walls from M-theory and string theory,
  Nucl. Phys. B {\bf 543}, 321 (1999),
  \href{http://www.arxiv.org/abs/hep-th/9807051}{hep-th/9807051};
  M. Cvetic, James T. Liu, H. Lu, C.N. Pope,
  Domain-wall Supergravities from Sphere Reduction
  Nucl. Phys. B {\bf 560}, 230 (1999),
  \href{http://www.arxiv.org/abs/hep-th/9905096}{hep-th/9905096}.

 \bibitem{str22}
  T. Shiromizu, D. Ida, H. Ochiai and T. Torii,
  Stability of $AdS_p \times S^n \times S^{q-n}$ Compactifications,
  Phys. Rev. D {\bf 64}, 084025 (2001),
  \href{http://www.arxiv.org/abs/hep-th/0106265}{hep-th/0106265};
  T. Torii, and T. Shiromizu,
  Cosmological constant, dilaton field and Freund-Rubin
  compactification,
  Phys. Lett. B {\bf 551}, 161 (2003),
  \href{http://www.arxiv.org/abs/hep-th/0210002}{hep-th/0210002}.






 \bibitem{tors2}
  F.W. Hehl, P. Heide, G.D. Kerlick and J.M. Nester,
  General relativity with spin and torsion: Foundations and
  prospects,
  Rev. Mod. Phys. {\bf 48}, 393 (1976).

 \bibitem{tors4}
  V. N. Ponomariov, A. Barvinsky and Yu. N. Obukhov,
  \emph{Geometrodynamical Methods and the Gauge Approach to the Gravitational Interactions}
  (Energoatomizdat, Moscow, 1985).

 \bibitem{tors5}
  B. Sazdovi\'{c},
  Torsion and nonmetricity in the stringly geometry,
  preprint
  \href{http://www.arxiv.org/abs/hep-th/0304086}{hep-th/0304086}.

 \bibitem{tors6}
  A.V. Minkevich and F. Karakura,
  On the relativistic dynamics of spinning mater in space-time
  with curvature and torsion,
  J. Phys. A {\bf 16}, 1409 (1983).

 \bibitem{tors7}
  G. Aprea, G. Montani, R. Ruffini,
  Test particles behavior in the framework of a lagrangian geometric theory
  with propagating torsion,
  Int. J. Mod. Phys. D {\bf 12}, 1875 (2003),
  \href{http://www.arxiv.org/abs/gr-qc/0401054}{gr-qc/0401054};
  S. M. Carroll and G. B. Field,
  Consequences of Propagating Torsion in Connection-Dynamic Theories of Gravity
  Phys. Rev. D {\bf 50}, 3867 (1994),
  \href{http://www.arxiv.org/abs/gr-qc/9403058}{gr-qc/9403058}.

 \bibitem{tors8}
  C. G. Boehmer,
  The Einstein static universe with torsion and the sign problem of the cosmological
  constant,
  Class. Quant. Grav. {\bf 21}, 1119 (2004),
  \href{http://www.arxiv.org/abs/gr-qc/0310058}{gr-qc/0310058}.






 \bibitem{br0}
  T. Kaluza,
  Zum Unit\"{a}tsproblem der Physik,
  Sitz. Preuss. Akad. Wiss. Phys. Math. K {\bf 1}, 966 (1921);
  O. Klein,
  The atomicity of electricity as a quantum theory law,
  Nature {\bf 118}, 516 (1926).

 \bibitem{br1a}
  T. Shiromizu, K. Maeda, and M. Sasaki,
  The Einstein equations on the 3-Brane World,
  Phys. Rev. D {\bf 62}, 024012 (2000),
  \href{http://www.arxiv.org/abs/gr-qc/9910076}{gr-qc/9910076};
  M. Sasaki, T. Shiromizu, and K. Maeda,
  Gravity, Stability and Energy Conservation on the Randall-Sundrum Brane-World,
  Phys. Rev. D {\bf 62}, 024008 (2000),
  \href{http://www.arxiv.org/abs/hep-th/9912233}{hep-th/9912233}.

 \bibitem{br1b}
  E. Anderson and R. Tavakol,
  Reformulation and Interpretation of SMS Braneworld,
  Class. Quant. Grav. {\bf 20}, L267 (2003),
  \href{http://www.arxiv.org/abs/gr-qc/0305013}{gr-qc/0305013}.

 \bibitem{br2}
  B. Mukhopadhyaya, S. Sen, S. Sen, S. SenGupta,
  Bulk Kalb-Ramond field in Randall Sundrum scenario,
  Phys. Rev. D {\bf 70}, 066009 (2004),
  \href{http://www.arxiv.org/abs/hep-th/0403098}{hep-th/0403098};
  B. Mukhopadhyaya, S. Sen, S. SenGupta,
  Does a Randall-Sundrum scenario create the illusion of a torsion-free
  universe?,
  Phys. Rev. Lett. {\bf 89}, 121101 (2002),
  \href{http://www.arxiv.org/abs/hep-th/0204242}{hep-th/0204242}.

 \bibitem{br3}
  O. Lebedev,
  Torsion Constraints in the Randall--Sundrum Scenario,
  Phys. Rev. D {\bf 65}, 124008 (2002),
  \href{http://www.arxiv.org/abs/hep-ph/0201125}{hep-ph/0201125};
  L. N. Chang, O. Lebedev, W. Loinaz, T. Takeuchi,
  Universal Torsion-Induced Interaction from Large Extra
  Dimensions,
  Phys. Rev. Lett. {\bf 85}, 3765 (2000), \newline
  \href{http://www.arxiv.org/abs/hep-ph/0005236}{hep-ph/0005236}.

 \bibitem{br4}
  A. Mennim and R. A. Battye,
  Cosmological expansion on a dilatonic brane-world,
  Class. Quant. Grav. {\bf 18}, 2171 (2001),
  \href{http://www.arxiv.org/abs/hep-th/0008192}{hep-th/0008192};
  C. Barcelo and M. Visser,
  Braneworld gravity: Influence of the moduli fields,
  J. High Energy Phys. {\bf 10}, 019 (2000),
  \href{http://www.arxiv.org/abs/hep-th/0009032}{hep-th/0009032}.

 \bibitem{br5}
  D. Gonta,
  AdSS$_5$ Brane World Cosmology,
  J. High Energy Phys. {\bf 12}, 007 (2004), \newline
  \href{http://www.arxiv.org/abs/hep-ph/0401117}{hep-ph/0401117}.

 \bibitem{br6}
  L. Randall and R. Sundrum,
  A Large Mass Hierarchy from a Small Extra Dimension,
  Phys. Rev. Lett. {\bf 83}, 3370 (1999),
  \href{http://www.arxiv.org/abs/hep-ph/9905221}{hep-ph/9905221};
  L. Randall and R. Sundrum,
  An Alternative to Compactification,
  Phys. Rev. Lett. {\bf 83}, 4690 (1999),
  \href{http://www.arxiv.org/abs/hep-th/9906064}{hep-th/9906064}.

 \bibitem{br7}
 T. Shiromizu, K. Koyama, S. Onda, and T. Torii,
 Can We Live On A D-Brane? Effective Theory on a Selfgravitating
 D-Brane,
 Phys. Rev. D {\bf 68}, 063506 (2003),
 \href{http://www.arxiv.org/abs/hep-th/0305253}{hep-th/0305253}.





 \bibitem{misc1}
  S. W. Hawking and C. F. R. Ellis,
  \emph{The Large Scale Structure of Spacetime},
  (Cambridge University Press, 1973).

 \bibitem{misc2}
  W. Israel,
  Singular Hypersurfaces and Thin Shells in General Relativity,
  Nuovo Cimento B {\bf 44}, 1 (1966).

 \bibitem{misc3}
  R. M. Wald,
  \emph{General Relativity},
  (Chicago University Press, Chicago, 1984).


\end{thebibliography}
\end{document}